\documentclass[final,5p,times,twocolumn]{elsarticle}

\usepackage{amssymb, amsmath, bm, graphicx, subcaption, lineno, mathtools}
\usepackage[numbers]{natbib}  

\graphicspath{{./plots/images}}

\journal{Journal of Computational Physics}

\begin{document}

\begin{frontmatter}

\title{Towards hybrid kinetic/drift-kinetic simulations in 6d Vlasov codes}

\author[ipp,tum]{M. Pelkner\corref{cor1}} 
\author[ipp]{K. Hallatschek} 
\author[ipp]{M. Raeth} 

\cortext[cor1]{Corresponding author. E-mail: maximilian.pelkner@ipp.mpg.de}

\affiliation[ipp]{organization={Max Planck Institute for Plasma Physics},
            addressline={Boltzmannstraße 2}, 
            city={Garching},
            postcode={85748}, 
            country={Germany}}

\affiliation[tum]{organization={TUM School of Natural Sciences},
            addressline={Boltzmannstraße 10}, 
            city={Garching},
            postcode={85748}, 
            country={Germany}}

\begin{abstract}
    \label{sec:abstract}

    Simulating fully kinetic, two-species plasmas is computationally challenging due to the stiff multiscale dynamics of electrons and ions. While enforcing a quasi-neutral time evolution mitigates this stiffness, it requires an electric potential that consistently maintains this constraint. In this work, we present an implicit approach to determine this electric field self-consistently within the semi-Lagrangian, fully kinetic BSL6D code. We employ a hybrid two-species model that couples kinetic ions with massless, drift-kinetic electrons, enabling an implicit treatment of the latter. Notably, the model captures the generation of ion-scale zonal flows. Beyond the algorithmic description, we provide a proof of second-order time-splitting error convergence under specific regularity assumptions. A key feature of our approach is an error-balancing mechanism: we demonstrate that the field solver achieves the required accuracy of the electric field by automatically adjusting the error of certain moments of the distribution function. Furthermore, we provide a comprehensive analysis of semi-Lagrangian interpolation errors to ensure robustness against the steep density and temperature gradients characteristic of tokamak edge plasmas. 

\end{abstract}
\end{frontmatter}

\section{Introduction}
\label{sec:introduction}

The study of turbulent transport remains a cornerstone of magnetic confinement fusion research, as it constitutes the primary mechanism for energy and particle loss in tokamaks \cite{wesson}. Over the past decades, gyrokinetic codes such as GENE \cite{gene}, GYRO \cite{gyro}, and ORB5 \cite{ORB5} have established themselves as the gold standard for simulating core plasma turbulence. However, experimental evidence indicates that critical confinement transitions, such as the L-H transition \cite{hMode} and the Greenwald density limit \cite{greenwald2007}, are governed by edge phenomena. In the plasma edge, where gradients and fluctuations are large, the gyrokinetic ordering reaches its limits, indicating that a transition to fully kinetic models is necessary. 

While fully kinetic simulations avoid the ordering uncertainties of the plasma edge, they have historically been considered computationally out of reach. It is only within the last decade that codes like GEMPIC \cite{gempic}, ssV \cite{SSV}, and BSL6D \cite{kormann2019} have demonstrated the feasibility of full-f kinetic simulations. Recent studies using BSL6D have provided first physically significant insights by demonstrating that non-gyrokinetic slab ion Bernstein wave turbulence can be excited under tokamak edge-compatible conditions \cite{raeth2023, bsl6dPRL2024}. Despite this success, a critical bottleneck remains: the physical scope of BSL6D is currently constrained by the assumption of adiabatic electrons and electrostaticity. To explore realistic fusion scenarios, a more sophisticated yet computationally efficient electron model is required.

A natural candidate is the drift-kinetic electron model, which captures the effective electron motion on ion timescales \cite{goldston} while reducing dimensionality. However, a direct coupling of kinetic ions and drift-kinetic electrons via the Vlasov-Poisson system introduces severe numerical stiffness, which arises from high-frequency "fast waves", such as the electron Langmuir and upper hybrid modes. Since explicit semi-Lagrangian schemes must resolve all physical scales to ensure numerical stability, these modes impose an extremely restrictive Courant-Friedrichs-Lewy (CFL) condition. This prevents the time step from reaching ion-scale durations necessary for turbulence studies, effectively nullifying the advantages of the hybrid approach.

To circumvent this multiscale constraint, the high-frequency dynamics must be eliminated by enforcing quasi-neutrality ($n_i = n_e$) at all times. The central numerical challenge lies in determining an electric potential that self-consistently maintains this constraint. In gyrokinetic hybrid models, this task is more tractable because the ion polarization term appears explicitly in the quasi-neutrality equation, providing a direct mathematical link to the potential. In a fully kinetic framework, however, the ion polarization is not explicitly available, but implicitly contained in the distribution function. This lack of an explicit polarization term constitutes a fundamental algorithmic hurdle, necessitating a novel approach to field calculation.

In this work, we present an implicit field solver that overcomes this challenge by deriving the quasi-neutral electric potential directly from the semi-Lagrangian advection algorithm of BSL6D. We restrict our analysis to the massless limit of drift-kinetic electrons, which enables an implicit treatment of the electron species and thereby reduces the complexity of our numerical analysis. The model is physically relevant as it captures the generation of zonal flows in the kinetic ion species accurately \cite{zonalFlow}. 

In the following, we first provide a brief overview of important aspects of the BSL6D code. Subsequently, we derive the algorithmic description of our field solver and provide a rigorous proof of second-order time-splitting error  convergence. A key aspect of our approach is an automatic error-balancing mechanism: we demonstrate that the field solver reaches the theoretical limit of the time-splitting accuracy by automatically adjusting the density and current density error of the distribution function. Furthermore, we provide a comprehensive analysis and correction of interpolation errors to ensure robustness of the scheme in the presence of steep gradients. Finally, we present numerical experiments to illustrate our findings. As our goal is to study tokamak edge physics, where significant density and temperature variations occur, we demonstrate that our method is also reliable in these parameter regimes. This work serves as the foundational step toward a full-f, two-species drift-kinetic electron model for edge turbulence research.

\section{BSL6D advection algorithm}
\label{sec:bsl6d advection algorithm}

\subsection{The semi-Lagrangian method}
\label{sec:semi-lagrangian method}

BSL6D solves the collisionless Vlasov equation for the ions,
\begin{equation}
    \label{vlasovEq}
    \partial_t f + \bm{v} \cdot \nabla f + \frac{e}{m_i} \left ( \bm{E} + \bm{v} \times \bm{B} \right ) \cdot \nabla_{\bm{v}} f = 0,
\end{equation}
via the semi-Lagrangian method \cite{kormann2019, schild2024}. This method exploits the fact that along certain trajectories in phase space, defined by 
\begin{align}
	  \label{characteristicX}
		\frac{d}{dt}\bm{X}(t) &= \bm{V}(t), \\
      \label{characteristicV}
		\frac{d}{dt}\bm{V}(t) &= \frac{e}{m_i} \bm{E}(\bm{X},t) + \frac{e}{m_i} \bm{V}(t) \times \bm{B},  
\end{align}
the distribution function remains constant, i.e., $df/dt = 0$. These trajectories, called characteristics, cannot intersect if the electric and magnetic field is at least Lipschitz continuous. Therefore, at any time $t$, every point $(\bm{x},\bm{v})$ in phase space lies on a unique characteristic $(\bm{X}(\bm{x}, \bm{v},t), \bm{V}(\bm{x}, \bm{v},t))$. This implies that any value $f(\bm{x},\bm{v},t)$ can be reconstructed from the initial condition $f(\bm{x}, \bm{v},t=0)$ via
\begin{equation}
    \label{backwardLagrange}
        f(\bm{x},\bm{v},t) = f(\bm{X}^{-1}(\bm{x},\bm{v},t), \bm{V}^{-1}(\bm{x},\bm{v},t), t = 0),
\end{equation}
where $\bm{X}^{-1}$ and $\bm{V}^{-1}$ are the backward characteristics satisfying
\begin{equation}
    \bm{X}^{-1}(\bm{X}(\bm{x},\bm{v},t), \bm{V}(\bm{x},\bm{v},t),t) = \bm{x}
\end{equation}
and 
\begin{equation}
   \bm{V}^{-1}(\bm{X}(\bm{x},\bm{v},t),\bm{V}(\bm{x},\bm{v},t),t)  = \bm{v}.
\end{equation}
BSL6D approximates the advection along the exact characteristics, which are generally unknown, using the algorithm described in section \ref{sec:bsl6d algorithm}.

\subsection{Rotating grid}
\label{sec:rotating grid}

In its current stage, BSL6D simulates an electrostatic, kinetic ion, adiabatic electron two-species model in a uniform magnetic background field. Since the gyrofrequency is a constant, numerical diffusion arising from interpolation errors can be reduced by transforming to a new coordinate system $(\bm{x}, \bm{v}) \rightarrow (\bm{x}, \bm{\tilde{v}})$, in which the velocity space frame rotates relative to the position space frame at precisely the gyrofrequency. This eliminates the gyration in velocity space \cite{sonnendrücker2004, schild2024convergence}. In the following, quantities expressed in the rotating frame are indicated with a tilde. Vectors transform as
\begin{equation}
    \label{vecTrafo}
    \bm{A}(t) = \bm{R}(t) \bm{\tilde{A}}, \qquad \bm{\tilde{A}}(t) = \bm{R}^{-1}(t) \bm{A},
\end{equation}
where the rotation matrix $\bm{R}$ is defined as
\begin{equation}
    \label{rotMat}
    \bm{R}(t) \coloneq \begin{pmatrix}
                \cos \Omega t & \sin \Omega t & 0\\
                - \sin \Omega t & \cos \Omega t & 0 \\
                0 & 0 & 1
                \end{pmatrix}, \qquad \Omega = \frac{e B}{m_i} .
\end{equation}
Note that the transformation (\ref{vecTrafo}) introduces a time dependency, e.g. $\bm{v}(t) =\bm{R}(t) \bm{\tilde{v}}$. For later reference, we also provide the matrix transformation law,
\begin{equation}
    \label{matTrafo}
    \bm{\Pi}(t) = \bm{R}(t) \bm{\tilde{\Pi}}\bm{R}^{-1}(t), \qquad \bm{\tilde{\Pi}}(t) = \bm{R}^{-1}(t) \bm{\Pi} \bm{R}(t) .
\end{equation}
In the new coordinates, Eqs. (\ref{characteristicX}) and (\ref{characteristicV}) take the form
\begin{align}
      \label{charactcharacteristicXRot}
        \frac{d}{dt}\bm{X}(t) &= \bm{V}(t), \\
	  \label{charactcharacteristicVRot}
        \frac{d}{dt}\bm{\tilde{V}}(t) &=  \frac{e}{m_i} \bm{\tilde{E}}(\bm{X},t)
\end{align}
and the Vlasov equation (\ref{vlasovEq}) for $f(\bm{x}, \bm{\tilde{v}},t)$ reads
\begin{equation}
      \label{vlasovEqRot}
        \partial_t f + \bm{v}(t) \cdot \nabla f + \frac{e}{m_i} \bm{\tilde{E}}\cdot \nabla_{\bm{\tilde{v}}} f = 0.
\end{equation}

\subsection{BSL6D algorithm} 
\label{sec:bsl6d algorithm}

\begin{figure}[t]
    \centering
    \includegraphics[width=0.6 \columnwidth]{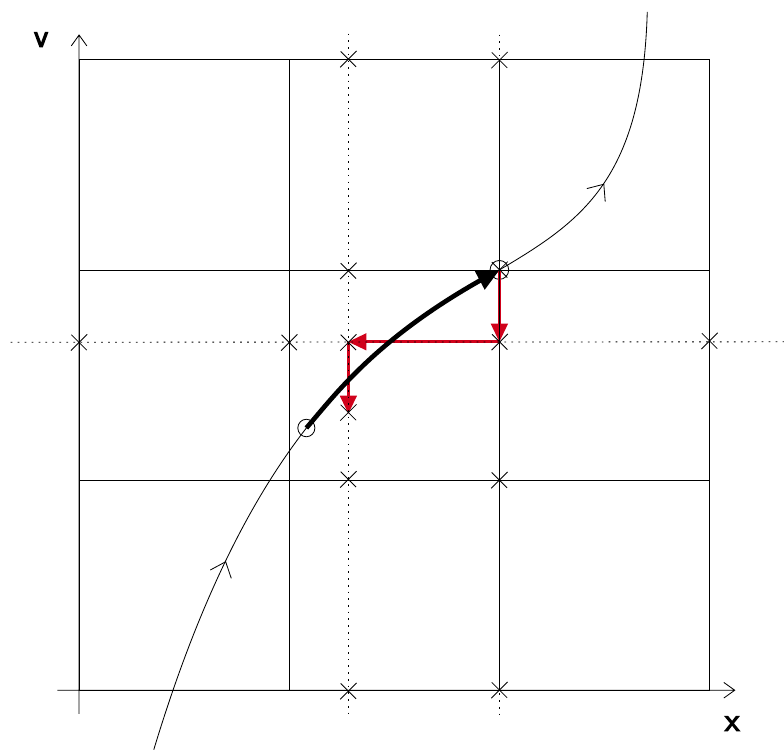} 
    \caption{Illustration of the semi-Lagrangian method on a 1D1V phase-space grid following Cheng and Knorr \cite{cheng1976}. To update the distribution function at a target grid node (top), the characteristic trajectory is back-traced over a time interval $\Delta t$. While exact characteristics are generally analytically inaccessible, they are approximated here via a Strang-splitting integration scheme (solid red arrows). The value at the foot of the trajectory (bottom) is determined using Lagrange interpolation and assigned to the target node. For computational efficiency, BSL6D decomposes multi-dimensional advections into a series of successive one-dimensional shifts (dashed lines). For further implementation details, see \cite{schild2024}.}
    \label{fig:strang_splitting}
\end{figure}

From now on, we set $m_i=e=B=1$, corresponding to coordinates in which time is measured in inverse ion gyrofrequencies, and introduce $\Delta t$ as the simulation time step. Let $t=0$ at the beginning of each step and introduce the subscript "$s$" to indicate quantities calculated at the beginning of step "$s$". The physical time $T$ is then defined by $T = s \Delta t$. As illustrated in Fig. \ref{fig:strang_splitting}, a single BSL6D step updates the distribution function $f_s$ on the 6D Cartesian grid as follows:
\begin{enumerate}
    \item First half v-advection: For all spatial positions, the distribution function is shifted in velocity space by 
    \begin{equation}
        \label{vShift}
        \Delta \bm{v} = \int_0^{\frac{1}{2} \Delta t} dt \bm{\tilde{E}}_s(T + t),
    \end{equation}
    where $\bm{\tilde{E}}_s(T + t)$ is defined by Eq. (\ref{vecTrafo}) and $\Delta \bm{v} = (\Delta v_x, \Delta v_y, \Delta v_z)$. The three-dimensional translation operator in velocity space, $\mathcal{T}_{\bm{v}}$, is approximated by three consecutive, one-dimensional Lagrange interpolations in each velocity direction,
    \begin{equation}
        \label{vInterpolation}
        f_s^* \coloneq \mathcal{I}_{\Delta v_x} \circ \mathcal{I}_{\Delta v_y} \circ \mathcal{I}_{\Delta v_z}[f_s].
    \end{equation}
    The details of the Lagrange interpolator $\mathcal{I}$ are discussed in Section \ref{sec:interpolation error correction}. Quantities calculated after this step are indicated by the superscript "$*$".
    \item x-advection: For all velocities, the distribution function is shifted in position space by 
    \begin{equation}
        \label{xShift}
        \Delta \bm{x} = \int_0^{\Delta t} dt \bm{v}(T + t),
    \end{equation}
    where $\bm{v}(T + t)$ is defined by Eq. (\ref{vecTrafo}) and $\Delta \bm{x} = (\Delta x, \Delta y, \Delta z)$. Note that while $\bm{\tilde{v}}$ is constant, $\bm{v}(T + t)$ evolves due to rotation. Analogous to the half v-advection step, the three-dimensional translation operator in position space, $\mathcal{T}_{\bm{x}}$, is approximated by three consecutive, one-dimensional Lagrange interpolations in each spatial direction,
    \begin{equation}
        \label{xInterpolation}
        f_s^{\dagger} \coloneq \mathcal{I}_{\Delta x} \circ \mathcal{I}_{\Delta y} \circ \mathcal{I}_{\Delta z} [f_s^*].
    \end{equation}
    Quantities determined after this step will be indicated by a superscript "$\dagger$".
    \item Field update: In the current BSL6D implementation with adiabatic electrons, the field update reads 
    \begin{equation}
        \phi_{s+1} = -\frac{T_e}{q n_{\text{bg}}}\ln(n_s^\dagger),
    \end{equation}
    where $T_e$ is the electron temperature and $n_{\text{bg}}$ is the background density. Throughout this work, $n_{\text{bg}} = 1$. The novel quasi-neutral field solver derived in Section \ref{sec:the electrostatic, quasi-neutral field solver} replaces this step.
    \item Final half v-advection: A second half v-advection step is performed using the updated electric field $\bm{\tilde{E}}_{s+1}$,
    \begin{equation}
        f_{s+1} \coloneq \mathcal{I}_{\Delta v_x} \circ \mathcal{I}_{\Delta v_y} \circ \mathcal{I}_{\Delta v_z} [f_s^{\dagger}].
    \end{equation}
\end{enumerate}
For technical reasons, the time step $\Delta t$ must be chosen so that the maximal shift in each step does not exceed one grid spacing \cite{schild2024}. Table \ref{tab:indices} summarizes the nomenclature.
\begin{table}[ht]
    \centering
    \begin{tabular}{|l|c|r|}
        \hline
        \textbf{advection step} & \textbf{index start} & \textbf{index end} \\
        \hline
        & & \\
        first $\tfrac{v}{2}$-adv. & $f_s$ & $f_s^*$ \\
        $x$-advection & $f_s^*$ & $f_s^\dagger$ \\
        second $\tfrac{v}{2}$-adv. & $f_s^\dagger$ & $f_{s+1}$ \\
        & & \\
        \hline
    \end{tabular}
    \caption{Overview of the indices used to indicate different stages of the numerical advection.}
    \label{tab:indices}
\end{table}

\section{The electrostatic, quasi-neutral field solver}
\label{sec:the electrostatic, quasi-neutral field solver}

\subsection{Massless limit of drift-kinetic electrons} 
\label{sec:massless limit of drift-kinetic electrons}

Assuming a uniform and constant background magnetic field in the $\hat{z}$ direction, the evolution of the collisionless, electrostatic, drift-kinetic electron distribution function in slab geometry is described by 
\begin{equation}
    \label{driftKineticElectrons}
    \partial_t f + \frac{\bm{E}\times \bm{B}}{B^2} \cdot \nabla f + v_z  \partial_z f - \frac{e}{m_e} E_z \partial_{v_z} f = 0.
\end{equation}
A key characteristic of realistic tokamak scenarios is the rapid electron motion along magnetic field lines, which allows them to efficiently neutralize electric potential perturbations within magnetic flux-surfaces. Such surfaces are defined by an irrational safety factor, i.e. a single magnetic field line ergodically covers the whole surface. In the following, we assume that all magnetic field lines trace out flux surfaces, that the slab analogs of these surfaces lie parallel to the y-z plane and that the electrons are massless. In the absence of collisions, these assumptions imply that electrostatic perturbations within a flux surface are instantaneously shielded, rendering the electron response adiabatic. Conversely, the flux-surface averaged electron density $\langle n_e \rangle$, where
\begin{equation}
    \label{fsOperation}
    \langle \cdot \rangle \coloneq \frac{1}{L_y L_z} \int dy dz \: \cdot \: ,
\end{equation}
remains constant because the flux-surface average of the $\bm{E} \times \bm{B}$ drift vanishes due to the periodic boundary conditions,
\begin{equation}
	\label{fixedEBackground}
		\partial_t \langle n_e \rangle = \langle \partial_y \phi \partial_x n_e - \partial_x \phi \partial_y n_e\rangle = 0.
\end{equation}
Consequently, electric potential perturbations with a non-vanishing flux-surface average cannot be shielded, but instead drive Langmuir oscillations in the flux-surface averaged ion density. To eliminate these high-frequency modes, we decompose the electric potential into its flux-surface average and a fluctuating part,
\begin{equation}
    \label{cfsaPotentiala}
		\phi = \langle \phi \rangle + (\phi - \langle \phi \rangle),
\end{equation}
and enforce the Boltzmann relation for the fluctuating component,
\begin{equation}
    \label{adiabaticPotential}
    \phi - \langle \phi \rangle = \frac{T_e}{e} \ln \left (\frac{n_e}{n_\text{bg}} \right ) - \frac{T_e}{e} \left \langle \ln \left (\frac{n_e}{n_\text{bg}} \right ) \right \rangle .
\end{equation}
Here, $T_e$ is the electron temperature, $e$ is the elementary charge and $n_{\text{bg}}=1$ is the background density \cite{Knorr:1970}. By further imposing quasi-neutrality, Eq. (\ref{cfsaPotentiala}) becomes
\begin{equation}
    \label{cfsaPotentialb}
		\phi = \langle \phi \rangle +  \frac{T_e}{e} \left (\ln(n) - \langle \ln (n) \rangle \right ),
\end{equation}
where $\langle \phi \rangle$ will be implicitly determined such that the quasi-neutrality constraint,
\begin{equation}
    \label{quasiNeutralityFSA}
    \langle n \rangle = \langle n_e \rangle,
\end{equation}
is rigorously maintained. Importantly, this approach eliminates the Langmuir oscillations in $\langle n \rangle$. The advantage of this model is that, although BSL6D strictly employs periodic boundary conditions (corresponding to rational q-profiles), the electron physics of irrational surfaces is analytically emulated. Consequently, while certain ion-scale geometric effects are excluded, the adiabaticity of electrons within flux surfaces is rigorously preserved. Crucially, the model can simulate the generation of slab zonal flows in the ion density, as $\partial_x \langle \phi \rangle \times \bm{B}$ flows within flux surfaces cannot be shielded by adiabatic electrons \cite{zonalFlow}.

\subsection{Calculating $\langle \phi \rangle$}
\label{sec:calculating <phi>}

To simplify the calculation of $\langle \phi \rangle$, we adopt a semi-discrete framework where only the time variable is discretized. Spatial variables are treated as continuous, effectively eliminating interpolation errors. We will discuss their role in the next section.

The global time discretization error of the current BSL6D scheme is $O(\Delta t^2)$ \cite{schild2024convergence}. To maintain this global convergence rate with the new field solver, our approach is to determine the electric potential such that the quasi-neutrality constraint is satisfied up to local $O(\Delta t^3)$ errors. Consequently, we first derive the evolution of the density $\langle n \rangle$ over a single simulation step, retaining terms up to $O(\Delta t^2)$. Subsequently, we impose the quasi-neutrality constraint to implicitly solve for $\langle \phi \rangle$.

A priori, it is not guaranteed that this implicitly determined electric potential maintains the global $O(\Delta t^2)$ accuracy of the complete distribution function. To rigorously establish this property, a formal convergence proof is provided at the end of this section.

\subsubsection{Time evolution of $\langle n \rangle$} 

While advecting the distribution function along the Strang-splitting trajectories $\bm{X}_{\text{split}}$ (constant $\bm{\tilde{v}}$) and $\bm{V}_{\text{split}}$ (constant $\bm{x}$), defined by 
\begin{equation}
    \bm{X_{\text{split}}}(\bm{x},\bm{\tilde{v}},t) \coloneq \int_0^t dt' \bm{v}(t') + \bm{x} 
\end{equation}
and
\begin{equation}
    \bm{V_{\text{split}}}(\bm{x},\bm{\tilde{v}},t) \coloneq \int_0^t dt' \bm{R}(t)\bm{E}(\bm{x},0) + \bm{\tilde{v}},
\end{equation}
the BSL6D scheme effectively integrates the constant coefficient advection equations 
\begin{equation}
    \label{bsl6dAdvectionX}
    \partial_t f + \partial_t \bm{X_{\text{split}}}\cdot \nabla f = 0 
\end{equation}
and 
\begin{equation}
    \label{bsl6dAdvectionV}
    \partial_t f + \partial_t \bm{V_{\text{split}}}\cdot \nabla_{\bm{\tilde{v}}} f = 0 .
\end{equation}
Note that $f$ in Eqs. (\ref{bsl6dAdvectionX}) and (\ref{bsl6dAdvectionV}) represents the numerical distribution function in the idealized limit of infinite spatial resolution, rather than the exact solution of the physical Vlasov equation. The particle density is defined by the first moment of the distribution function,
\begin{equation}
    \label{density}
    n(\bm{x},t)\coloneq \int d^3 \tilde{v} f(\bm{x}, \tilde{\bm{v}},t),
\end{equation}
and remains invariant during v-advection steps according to Eq. (\ref{bsl6dAdvectionV}),
\begin{equation}
    0 = \partial_t \int d^3 \tilde{v} f +  \int d^3 \tilde{v} \bm{\tilde{E}} \cdot \nabla_{\bm{\tilde{v}}} f = \partial_t n .
\end{equation}
Conversely, integrating Eq. (\ref{bsl6dAdvectionX}) over the velocity space gives the continuity equation,
\begin{equation}
    \label{zerothMomEq}
    \partial_t n = - \nabla \cdot \bm{j},
\end{equation}
where $\bm{j} = \bm{R}(t) \bm{\tilde{j}}$ denotes the current density,
\begin{equation}
    \label{currentDensity}
    \bm{\tilde{j}}(\bm{x}, t) \coloneq \int d^3 \tilde{v} \bm{\tilde{v}} f(\bm{x}, \bm{\tilde{v}},t).
\end{equation}
Integrating Eq. (\ref{zerothMomEq}) over the time interval $\Delta t$ and applying the flux-surface average operation (\ref{fsOperation}) provides the change in $\langle n \rangle$ over a Strang-splitting step,
\begin{equation}
    \label{nCh1}
    \langle n_{s+1} (\bm{x}) \rangle - \langle n_s (\bm{x})\rangle = \left \langle - \nabla \cdot \int_T^{T + \Delta t} dt \bm{j}(\bm{x},t) \right \rangle .
\end{equation}
By evaluating the flux-surface  operation (\ref{fsOperation}), Eq. (\ref{nCh1}) simplifies to
\begin{equation}
    \label{nCh2}
    \langle n_{s+1} \rangle - \langle n_s \rangle  = - \hat{x} \partial_x \cdot \int_T^{T + \Delta t} dt  \left \langle \bm{j}(t) \right \rangle .
\end{equation}
Since the advection equations (\ref{bsl6dAdvectionX}) and (\ref{bsl6dAdvectionV}) are integrated exactly in the adopted semi-discrete framework, we can expand the current density in Eq. (\ref{nCh2}),
\begin{align}
   \label{nCh3}
   &\langle n_{s+1} \rangle - \langle n_s \rangle = \notag \\
   & - \hat{x} \partial_x  \cdot \left (\int_T^{T + \Delta t} dt \left [ \left \langle \bm{j}_s^* (t)\right \rangle + \bm{R}(t) \int_T^{T + t} dt' \partial_{t'} \left \langle \bm{\tilde{j}}(t') \right \rangle \right ] \right ).
\end{align}
The leading current density term is evaluated at the onset of the x-advection (after the initial half v-advection). While this term is constant in the rotating frame, it introduces a time dependence in the non-rotating frame due to the coordinate transformation (\ref{vecTrafo}). The temporal evolution of the current density in Eq. (\ref{nCh3}) is governed by the first moment of Eq. (\ref{bsl6dAdvectionX}),
\begin{equation}
    \label{firstMomPre}
    \partial_t \bm{\tilde{j}} + \int d^3 \tilde{v} \bm{\tilde
    {v}} \bm{v} \cdot \nabla f = 0,
\end{equation}
which, by introducing the momentum flux density tensor,
\begin{equation}
    \label{momentumFluxDensity}
    \bm{\tilde{\Pi}}(\bm{x},t) \coloneq \int d^3 \tilde{v} \bm{\tilde{v}} \bm{\tilde{v}} f(\bm{x}, \bm{\tilde{v}},t), 
\end{equation}
becomes
\begin{equation}
    \label{firstMomEq}
    \partial_t \bm{\tilde{j}} = - \nabla \cdot \bm{R} \bm{\tilde{\Pi}} .
\end{equation}
Thus, Eq. (\ref{nCh3}) takes the form
\begin{align}
    \label{nCh4}
     &-\left(
     \langle n_{s+1} \rangle - \langle n_s \rangle \right) = \notag \\
     &\hat{x}\partial_x \cdot \int_T^{T + \Delta t} dt \left [ \left \langle \bm{j}_s^* (t)\right \rangle -  \partial_x \int_T^{T + t} dt' \bm{R}(t') \left \langle \bm{\tilde{\Pi}}(t') \right \rangle \bm{R}^{-1}(t) \cdot \hat{x} \right ] .  
\end{align}
Since the variation of the momentum flux density during the x-advection contributes only $O(\Delta t^3)$ to the density change, i.e., at the order of the inherent Strang-splitting error, it can be neglected. Therefore, 
\begin{align}
    \label{nCh5}
     &\langle n_{s+1} \rangle - \langle n_s \rangle = \notag \\
     & -\hat{x}\partial_x \cdot \int_T^{T + \Delta t} dt \left [ \left \langle \bm{j}_s^*(t) \right \rangle - \partial_x \int_T^{T + t} dt' \bm{R}(t') \left \langle \bm{\tilde{\Pi}}_s^* \right \rangle \bm{R}^{-1}(t) \cdot \hat{x} \right ] \notag \\
     &+ O(\Delta t^3).  
\end{align}
Shifting the integration variable of Eq. (\ref{nCh5}) by $\Delta t/2$ allows us to exploit the identity
\begin{equation}
    \label{c1}
    \int_{-\frac{\Delta t}{2}}^{\frac{\Delta t}{2}} dt \bm{R}(t) = c_1 \Delta t \mathbb{I} , \qquad \text{where} \qquad c_1 \coloneq \text{sinc} \left(\tfrac{\Omega \Delta t}{2} \right) ,
\end{equation}
because 
\begin{align}
    \label{jCh1}
    &-\hat{x} \partial_x \cdot \int_T^{T + \Delta t} dt \langle \bm{j}_s^* (t) \rangle\notag \\
    & \qquad \qquad = -\hat{x} \partial_x \cdot \int_0^{\Delta t} dt \bm{R}(T + t) \left \langle \bm{\tilde{j}}_s^* \right \rangle \notag \\ 
    & \qquad \qquad = -\hat{x} \partial_x \cdot \int_0^{\Delta t} dt \bm{R}\left (t - \tfrac{\Delta t}{2} \right ) \left \langle \bm{j}_s^* \left (T + \tfrac{\Delta t}{2} \right ) \right \rangle \notag \\ 
    & \qquad \qquad =-\hat{x} \partial_x \cdot \int_{-\frac{\Delta t}{2}}^{\frac{\Delta t}{2}} dt \bm{R}(t) \left \langle \bm{j}_s^* \left (T + \tfrac{\Delta t}{2} \right ) \right \rangle \notag \\
    &\qquad \qquad = - \partial_x  \left \langle (\bm{j}_s^*)_x \left (T + \tfrac{\Delta t}{2} \right ) \right \rangle c_1 \Delta t . 
\end{align}
To further streamline the momentum flux density tensor integral, we decompose it into its irreducible components,
\begin{equation}
    \label{matDecomposition}
    \left \langle \bm{\tilde{\Pi}}_s^* \right \rangle = a \underbrace{  
    \begin{pmatrix}
    1 & 0 \\
    0 & 1
    \end{pmatrix}}_{\hat{e}_1}
    + b
    \underbrace{\begin{pmatrix}
    1 & 0 \\
    0 & -1
    \end{pmatrix}}_{\hat{e}_2}
    + c
    \underbrace{\begin{pmatrix}
    0 & 1 \\
    1 & 0
    \end{pmatrix}}_{\hat{e}_3},
\end{equation}
where
\begin{equation}
    \label{decompositionCoefficients}
     a \coloneq \tfrac{1}{2} \mathrm{Tr} \left ( \left \langle \bm{\tilde{\Pi}}_s^* \right \rangle \right ), \:b \coloneq \tfrac{1}{2} \left [ \left \langle (\bm{\tilde{\Pi}}_s^*)_{xx} \right \rangle - \left \langle (\bm{\tilde{\Pi}}_s^*)_{yy} \right \rangle \right ], \: c \coloneq \left \langle (\bm{\tilde{\Pi}}_s^*)_{xy} \right \rangle .
\end{equation}
The matrices $\hat{e}_2$ and $\hat{e}_3$ form a basis of the two-dimensional vector space of symmetric and traceless $2 \times 2$ matrices. It can be shown that the traceless components $b$ and $c$ transform according to a Spin-2 representation of the $SO(2)$ group. That is, for a rotation $\phi = \Omega t$ of the velocity frame, the components transform as
\begin{equation}
    \label{spin2Representation}
    \begin{pmatrix}
    b \\
    c
    \end{pmatrix}
    \rightarrow 
    \begin{pmatrix}
    b'\\
    c'
    \end{pmatrix}
    =
    \begin{pmatrix}
    \cos(2 \Omega t) b + \sin(2 \Omega t) c \\
    - \sin(2 \Omega t) b  + \cos(2 \Omega t) c 
    \end{pmatrix} .
\end{equation}
Therefore, upon replacing
\begin{equation}
    \bm{R}(t') \left \langle \bm{\tilde{\Pi}}_s^* \right \rangle \bm{R}^{-1}(t) \rightarrow \bm{R}(t'-t) \left \langle \bm{\Pi}_s^*(t) \right \rangle 
\end{equation}
in Eq. (\ref{nCh5}), the momentum flux density integral gives
\begin{align}
     &\int_T^{T + \Delta t} dt \int_T^{T + t} dt' \bm{R}(t') \left \langle \bm{\tilde{\Pi}}_s^* \right \rangle \bm{R}^{-1}(t) \notag \\
     &= \int_0^{\Delta t} dt \int_0^{t} dt' \bm{R} (t'-t) \left [ a \hat{e}_1 + b'(T + t) \hat{e}_2 + c'(T + t) \hat{e}_3 \right ], 
\end{align}
from which it follows that
\begin{align}
    \label{intermediateA}
    &\hat{x} \cdot \int_T^{T + \Delta t} dt \int_T^{T + t} dt' \bm{R}(t') \left \langle \bm{\tilde{\Pi}}_s^* \right \rangle \bm{R}^{-1}(t) \cdot \hat{x} \notag \\
    & \qquad =\int_0^{\Delta t} dt \int_0^{t} dt' \big [ \cos(\Omega (t'-t)) \left (a + b'(T + t) \right ) \notag \\
    & \qquad \quad + \sin(\Omega(t'-t)) c'(T + t) \big ].
\end{align}
Evaluating this double integral yields
\begin{align}
    \label{intermediateB}
    &\hat{x} \cdot \int_T^{T + \Delta t} dt \int_T^{T + t} dt' \bm{R}(t') \left \langle \bm{\tilde{\Pi}}_s^* \right \rangle \bm{R}^{-1}(t) \cdot \hat{x} = \notag \\ 
    &\qquad \qquad \tfrac{1}{2} \Delta t^2 \text{sinc}\left(\tfrac{\Omega \Delta t}{2}\right)^2\left (a+ b'(T + \Delta t) \right ) .
\end{align}
On the other hand, Eqs. (\ref{matDecomposition}), (\ref{decompositionCoefficients}) and (\ref{spin2Representation}) imply that
\begin{equation}
\tfrac{1}{2} \Delta t^2 \text{sinc}\left(\tfrac{\Omega \Delta t}{2}\right)^2\left ( a + b'(T + \Delta t) \right ) = \tfrac{1}{2} \left \langle (\bm{\Pi}_s^*)_{xx} \left(T + \tfrac{\Delta t}{2}\right)\right \rangle c_1^2 \Delta t^2 .
\end{equation}
In summary, we obtain the compact expression
\begin{align}
    \label{nCh6}
    &\langle n_{s+1} \rangle - \langle n_s \rangle = \notag \\
    &- \partial_x \left \langle (\bm{j}^*_s)_x \left (T + \tfrac{\Delta t}{2} \right ) \right \rangle c_1 \Delta t + \tfrac{1}{2}\partial^2_x \left \langle (\bm{\Pi}_{s-1}^\dagger)_{xx} \left(T + \tfrac{\Delta t}{2}\right)\right \rangle c_1^2 \Delta t^2 \notag \\
    &+ O(\Delta t^3).
\end{align}
Note that we also neglect the change in the momentum flux density tensor during the v-advection step; thus, we use the index "$\dagger$" instead of "$*$". 

\subsubsection{Determining $\langle \phi \rangle$ via the quasi-neutrality constraint}

To obtain an expression for the current density at the beginning of the x-advection, $\langle (\bm{j}_s^*)_x(t) \rangle$, we integrate Eq. (\ref{bsl6dAdvectionV}) over velocity space,
\begin{equation}
    \label{jCh2}
    \left \langle \bm{\tilde{j}}_s^*\right \rangle = \left \langle \bm{\tilde{j}}_{s-1}^\dagger \right \rangle  + \int_{-\frac{\Delta t}{2}}^{+\frac{\Delta t}{2}} dt'\bm{R}^{-1}(T + t') \langle \bm{E}_s n_s \rangle .
\end{equation}
The flux-surface averaged electric field, $\langle \bm{E}_s \rangle$, can be extracted from the term $\langle \bm{E}_s n_s \rangle$ by substituting $\bm{E} = - \nabla \phi$,
\begin{align}
    \left \langle \bm{E}_s n_s \right \rangle &= \left \langle n_s \left (\bm{E}_s + \langle \bm{E}_s \rangle - \langle \bm{E}_s \rangle \right ) \right \rangle \notag \\
    &= - \left \langle \partial_x \phi_s \right \rangle  \left \langle n_s \right \rangle \hat{x} - \left \langle n_s \nabla \left (\phi_s - \langle \phi_s \rangle \right ) \right \rangle.
\end{align}
Since
\begin{equation}
    \label{jCh3}
    \bm{R} \left (T + \tfrac{\Delta t}{2}\right ) \int_{-\frac{\Delta t}{2}}^{\frac{\Delta t}{2}} dt \bm{R}^{-1}(T + t) = \Delta t
    \begin{pmatrix}
    c_2 & \frac{\Omega \Delta t}{2} c_1^2 & 0\\
    - \frac{\Omega \Delta t}{2} c_1^2 & c_2 & 0 \\
    0 & 0 & 1
\end{pmatrix} ,
\end{equation}
where $c_2 = \text{sinc}\left ( \Omega \Delta t \right )$, Eq. (\ref{jCh2}) gives
\begin{align}
    \label{jCh4}
   & \left \langle (\bm{j}_s^*)_x \left ( T + \tfrac{\Delta t}{2}\right ) \right \rangle = \left \langle (\bm{j}_{s-1}^\dagger)_x \left (T + \tfrac{\Delta t}{2} \right ) \right \rangle \notag \\
   & \qquad - \langle \partial_x \phi_s \rangle \langle n_s \rangle c_2 \Delta t  \langle \partial_x ( \phi_s - \langle \phi_s \rangle ) n_s \rangle c_2 \Delta t \notag \\
   & \qquad - \tfrac{1}{2} \langle \partial_y ( \phi_s - \langle \phi_s \rangle ) n_s \rangle c_1^2 \Omega \Delta t^2 .
\end{align}
Using the adiabatic electron response on flux surfaces in conjunction with quasi-neutrality (\ref{cfsaPotentialb}), 
\begin{equation}
    \phi - \langle \phi \rangle = \frac{T_e}{e} \left ( \ln(n) - \langle \ln(n) \rangle \right ) ,
\end{equation}
Eq. (\ref{jCh4}) becomes
\begin{align}
    \label{jCh5}
    \left \langle (\bm{j}_s^*)_x \left (T + \tfrac{\Delta t}{2} \right ) \right \rangle =& \left \langle (\bm{j}_{s-1}^\dagger)_x \left (T + \tfrac{\Delta t}{2} \right ) \right \rangle - \langle \partial_x  \phi_s \rangle \langle n_s \rangle c_2 \Delta t \notag \\
    &- \frac{T_e}{e} \left \langle \mathcal{D} \left (\ln (n_s) - \langle \ln(n_s) \rangle \right ) n_s \right \rangle ,
\end{align}
where
\begin{equation}
    \label{defD}
    \mathcal{D} \coloneq \left ( \partial_x  +  c_3 \partial_y \right ) \quad \text{and} \quad c_3 \coloneq \frac{c_1^2}{2 c_2} \Omega \Delta t  = \tan \left (\frac{\Omega \Delta t}{2} \right ) .
\end{equation}
Simultaneously, calculating the antiderivative of Eq. (\ref{nCh6}) with respect to $x$ provides an alternative expression for the current,
\begin{align}
    \label{jCh6}
    &\left \langle (\bm{j}_s^*)_x \left (T + \tfrac{\Delta t}{2} \right ) \right \rangle = -\frac{1}{c_1 \Delta t} \left (\int dx \langle n_{s+1} \rangle -  \langle n_s \rangle \right )+ \notag \\
    & \qquad \qquad \tfrac{1}{2} \partial_x \left \langle (\bm{\Pi}_{s-1}^\dagger)_{xx} \left(T + \tfrac{\Delta t}{2}\right)\right \rangle c_1 \Delta t + C + O(\Delta t^2). 
\end{align}
The appearing undetermined integration constant $C$ represents a spatially uniform, divergence-free background current. Assuming periodic boundary conditions and zero net current, this constant vanishes identically without loss of generality. By requiring the quasi-neutrality constraint
\begin{equation}
    \label{quasiNeutralityFSA2}
    \langle n_{s+1} \rangle = \langle n_e \rangle + O(\Delta t^3)
\end{equation}
to be strictly satisfied and equating the derived current density expressions  (\ref{jCh5}) and (\ref{jCh6}), we obtain the final governing equation for  $\partial_x \langle \phi_s \rangle$,
\begin{align}
    \label{phiXCFSApre}
    &\langle \partial_x \phi_s \rangle = \notag \\
    &\frac{1}{\langle n_s \rangle} \frac{1}{c_2 \Delta t} \left \langle (\bm{j}_{s-1}^\dagger)_x \left (T + \tfrac{\Delta t}{2} \right ) \right \rangle - \frac{1}{2 \langle n_s \rangle} \frac{c_1}{c_2} \partial_x \left \langle (\bm{\Pi}_{s-1}^\dagger)_{xx} \left(T + \tfrac{\Delta t}{2}\right)\right \rangle \notag \\
    & - \frac{1}{\langle n_s \rangle}\frac{1}{c_1 c_2 \Delta t^2} \int dx \left \langle n_s - n_e \right \rangle - \frac{T_e}{e \langle n_s \rangle} \langle n_s \mathcal{D} (\ln(n_s) - \langle \ln(n_s) \rangle) \rangle \notag \\
    & + O(\Delta t) . 
\end{align}
For the initial simulation step, this relation is replaced by its modified counterpart accounting for the half-step initialization,
\begin{align}
    \label{initPhiXCFSA}
    &\langle \partial_x \phi_0 \rangle = \notag \\
    &\frac{1}{\langle n_0 \rangle} \frac{2}{c_1 \Delta t} \left \langle (\bm{j}_0)_x \left (\tfrac{\Delta t}{2} \right ) \right \rangle - \frac{1}{\langle n_0 \rangle}\partial_x \left \langle (\bm{\Pi}_0)_{xx} \left(\tfrac{\Delta t}{2}\right)\right \rangle \notag \\
    & - \frac{1}{\langle n_0 \rangle}\frac{2}{c_1^2 \Delta t^2} \int dx \left \langle n_0 - n_e \right \rangle - \frac{T_e}{e \langle n_0 \rangle} \langle n_0\mathcal{D}_0 (\ln(n_0) - \langle \ln(n_0) \rangle) \rangle \notag \\
    &+ O(\Delta t) ,
\end{align}
with the modified operator
\begin{equation}
    \mathcal{D}_0 \coloneq \partial_x + \tan\left( \frac{\Omega \Delta t}{4}\right)\partial_y .
\end{equation}
In summary, the numerical electric potential is given by 
\begin{equation}
    \label{phiCFSA}
    \bm{E}_s \coloneq -\langle \partial_x \phi_s \rangle \hat{x} - \frac{T_e}{e} \nabla (\ln (n_s) - \langle \ln(n_s) \rangle) ,
\end{equation}
where 
\begin{align}
    \label{phiXCFSA}
    &\langle \partial_x \phi_s \rangle \coloneq \notag \\
    &\frac{1}{\langle n_s \rangle} \frac{1}{c_2 \Delta t} \left \langle (\bm{j}_{s-1}^\dagger)_x \left (T + \tfrac{\Delta t}{2} \right ) \right \rangle - \frac{1}{2 \langle n_s \rangle} \frac{c_1}{c_2} \partial_x \left \langle (\bm{\Pi}_{s-1}^\dagger)_{xx} \left(T + \tfrac{\Delta t}{2}\right)\right \rangle \notag \\
    & - \frac{1}{\langle n_s \rangle}\frac{1}{c_1 c_2 \Delta t^2} \int dx \left \langle n_s - n_e \right \rangle - \frac{T_e}{e \langle n_s \rangle} \langle n_s \mathcal{D} (\ln(n_s) - \langle \ln(n_s) \rangle) \rangle 
\end{align}
for $s>0$.

\subsubsection{Error convergence of the density}

\begin{figure}[!t] 
    \centering
    \begin{subfigure}[t]{0.75\columnwidth}
        \centering
        \includegraphics[width=\textwidth]{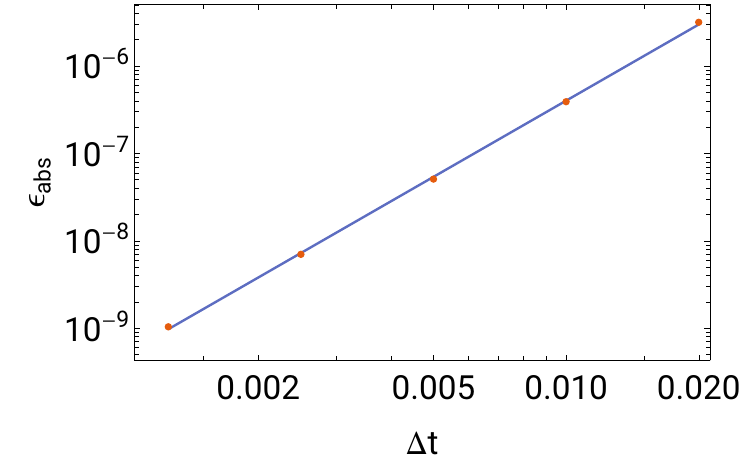}
        \caption{Dissipative density error convergence. The fitted line indicates a convergence order of $\Delta t^{2.9}$.}
        \label{fig:n_conv_abs}
        \vspace{0.5cm}
    \end{subfigure}
    \begin{subfigure}[t]{0.75\columnwidth}
        \centering
        \includegraphics[width=\textwidth]{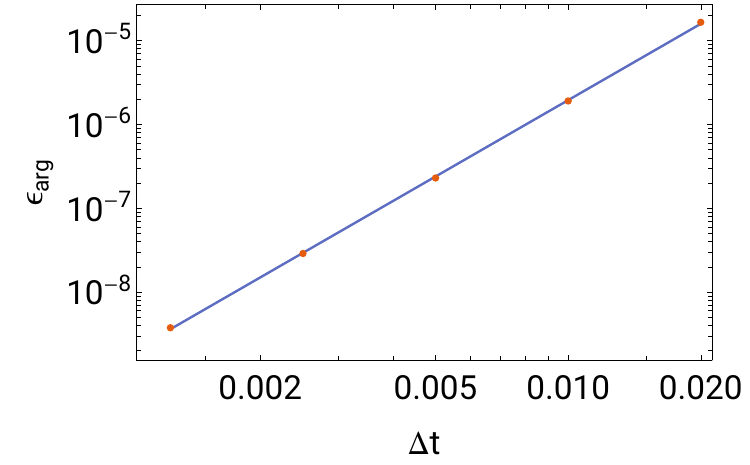}
        \caption{Dispersive density error convergence. The fitted line indicates a convergence order of $\Delta t^{3.0}$.}
        \label{fig:n_conv_arg}
    \end{subfigure}
    \caption{Illustration of the density error convergence order. The ion distribution function was initialized as $f = n_0 (1 + \alpha \delta n) f_{\text{M}}(v)$ with a constant electron background $n_e =  (1 + \alpha \delta n)$ on a 1D2V grid ($32 \times 65 \times 65$ points, domain $[0,2\pi] \times [-8,8]\times [-8,8]$). Here, $\delta n = \sin(x)$ denotes the background density variation with magnitude $\alpha = 0.01$. As there is only one spatial dimension in the x-direction, $\langle n \rangle = n$.}
    \label{fig:n_conv}
\end{figure}

To verify the theoretical $O(\Delta t^3)$ density error convergence of our scheme, we decompose the numerical error into its dissipative and dispersive components and consider a one-dimensional setup with the spatial direction oriented in the $\hat{x}$ direction (i.e., orthogonally to the flux-surfaces). Since this setup constitutes a stationary state where analytically $\partial_t \langle n \rangle = \partial_t n = 0$, we can use the density at $t=0$ as the exact reference solution. In Figure \ref{fig:n_conv}, we plot the two errors
\begin{equation}
    \epsilon_{\text{abs}} \coloneq 1 - \frac{|\hat{n}^{\text{num}}(k,t=25)|}{|\hat{n}^{\text{num}}(k,t=0)|} 
\end{equation}
and 
\begin{equation}
    \epsilon_{\text{arg}} \coloneq |\arg (\hat{n}^{\text{num}}(k,t=0)) - \arg (\hat{n}^{\text{num}}(k,t=25))| 
\end{equation}
for the wavenumber $k=1$ to minimize the influence of spatial interpolation errors.

\subsection{Interpolation error correction}
\label{sec:interpolation error correction}

Numerical errors arising from interpolation can manifest as unphysical density drifts or numerical instabilities, particularly in long-term simulations. In this section, we analyze these errors spectrally and introduce a correction scheme to ensure consistency between the advection operators and the field solver.

\subsubsection{BSL6D Lagrange interpolation}
\label{sec:lagrange interpolation}

For a one-dimensional spatial shift $d \in [0,1]$, the one-dimensional interpolation operator $\mathcal{I}_d[f]$ introduced in Section \ref{sec:bsl6d algorithm} is defined pointwise via
\begin{align}
    \label{lgInterpolation}
    \mathcal{I}_d[f(x_n)] \coloneq& \sum_{i = n - m}^{n + m} f(x_i) \prod_{j \neq i} \frac{(n + d) - j}{i - j} \notag \\
    =& \sum_{j=n-m}^{n+m} f(x_j) T_d(n-j),
\end{align}
where $m \coloneq (S-1)/2$ and $T_d$ is the translation kernel of Lagrange interpolation \cite{stoer}. The subsequent discussion focuses on odd "stencils" $S$, because current technical limitations of the BSL6D code do not permit error correction for even stencils. Given periodic boundary conditions, i.e., $f(x_N) = f(x_0)$, the discrete exponentials,
\begin{equation}
    \label{discreteExponential}
   e^{i \frac{2 \pi}{N} n k} = \left ( 1, e^{i \frac{2 \pi}{N} k }, e^{i \frac{2 \pi}{N} 2 k},\ldots , e^{i \frac{2 \pi}{N} (N-1) k} \right ), \quad n \in \{0, \ldots, N-1\},
\end{equation}
act as eigenvectors of the interpolator because
\begin{align}
    \mathcal{I}_d \left [ e^{i \frac{2 \pi}{N} n k} \right ] :=& \sum_{j=n-m}^{n+m} T_d(n-j) e^{i \frac{2 \pi}{N} j k} \notag \\
     =& \sum_{p = m}^{-m} T_d(p) e^{i \frac{2 \pi}{N} (n - p) k} = \delta(d) e^{i \frac{2 \pi}{N} n k} ,
\end{align}
where $\delta(d) \in \mathbb{C}$ is defined as
\begin{equation}
    \label{interpolationError}
    \delta(d) := \sum_{j = m}^{-m} T_d(j) e^{- i \frac{2 \pi}{N} j k}.
\end{equation}
An exact shift implies $\delta(d) = e^{i k d}$. Any deviation from the exact shift constitutes the interpolation error, which consists of dissipative (amplitude) and dispersive (phase) components. Interpolations in velocity space are defined analogously upon changing the weights $f(x_i) \rightarrow f(v_i)$ and adapting the calculation of the one-dimensional shifts $d$ as discussed in Sec. \ref{sec:bsl6d algorithm}.

\subsubsection{Moment conservation in velocity space}

A key property of the $S$-point Lagrange interpolation (\ref{lgInterpolation}) is the exact reproduction of moments up to order $S-1$.  Consequently, no error correction is needed for the velocity-space interpolations if $S \geq 3$, since Eq. (\ref{phiCFSA}) contains only moments up to second order. To prove this, recall that 
\begin{equation}
  \sum_{j=n-m}^{n+m} T_d (i-j) P(x_j) = P(x_i + d) 
\end{equation}
for any discrete polynomial $P$ of order less than $S$ \cite{stoer}. Therefore, any moment up to order $S-1$ of the $T$-shifted distribution function is equal to the $(-d)$-shifted moment of the initial distribution function, since
\begin{align}
  \sum_i i^m \sum_j T_d(i-j) f_j =& \sum_{j}f_j \sum_i T_d(i-j) i^m \notag \\
  =& \sum_{j}f_j \sum_k T_d(-k) (j-k)^m \notag \\
  =&\sum_{j}(j-d)^mf_j,
\end{align}
where $k=j-i$. We now turn to the more subtle errors arising from position-space interpolations.

\subsubsection{Position space interpolation error correction}

The spectral interpretation of Eq. (\ref{interpolationError}) is that the interpolation operation $\mathcal{I}_d[f]$ does not shift the discrete exponentials as expected from the exact shift operator $\mathcal{T}$, 
\begin{equation}
    \label{perfectShift}
    \mathcal{T}_d \left [ e^{i \frac{2 \pi}{N} n k} \right ] = e^{i k d}  e^{i \frac{2 \pi}{N} n k} = (1 + i k d - \tfrac{k^2}{2} d^2 + ...) e^{i \frac{2 \pi}{N} n k},
\end{equation}
but instead introduces wavenumber-dependent errors,
\begin{align}
  \label{interpolatedShift}
    \mathcal{I}_d \left [ e^{i \frac{2 \pi}{N} n k} \right ] =& \sum_{j=n-m}^{n+m} T_d(n-j) e^{i \frac{2 \pi}{N} j k} \notag \\
     =& (1 + i\alpha(k, S) d - \beta(k, S) d^2 + ...) e^{i \frac{2 \pi}{N} n k}.
\end{align}
These interpolation errors lead to an incorrect position space advection of the distribution function $f$,
\begin{equation}
    \label{distributionFunctionFou}
    f(x_n, v_j) = \frac{1}{N} \sum_{k = 0}^{N-1} \hat{f}_k (v_j) e^{i \frac{2 \pi}{N} n k} ,
\end{equation}
and therefore to an erroneous evolution of the discrete moments, in particular the density. Thus, without correction, Eq. (\ref{phiXCFSA}) calculates an electric potential that is erroneous on discrete grids. To derive an error correction, we first identify the modified wavenumbers $\alpha(k,S)$ and $\beta(k,S)$ from Eqs. (\ref{lgInterpolation}) and (\ref{interpolationError}). Starting with $\beta(k, S)$ for odd $S$, we can use the ansatz
\begin{equation}
  \beta(k,S)=\sum_{n=1}^{(S+1)/2}b_n \sin^{2n}\frac{k}2, \label{eq:betaansatz}
\end{equation}
because $\beta$ is symmetric in $k$, the power series (\ref{eq:betaansatz})
contains only integer wavenumbers corresponding to the integer grid points in real space, as 
\begin{equation}
    \sin^2(k/2)=(1-\cos k)/2,
\end{equation}
the sum is numerically well-behaved, as it contains only positive summands, and the condition of $L$ being an $S$-th order interpolation operator implies $\beta(k,S)-k^2/2$ has a zero of order $S$ at $k=0$. The coefficients $b_n$ are determined by substituting $x:=\sin(k/2)$ and noting $\beta(k,S)\rightarrow k^2/2$ for $S\rightarrow\infty$, i.e.
\begin{equation}
  \sum_{n=1}^{\infty}b_n x^{2n}=\frac{k^2}2=2\arcsin^2 x\quad\Rightarrow b_n
  =\frac{2^{2n-2}}{n^2\binom{2n}{n}},
\end{equation}
with the series for $\arcsin^2x$ taken from \cite{gradshteyn} 1.645.2. An analogous argument can be made for the function $\alpha$, which gives
\begin{align}
  \alpha(k,S) =& \partial_k\beta(k,S) \notag \\
  =& \cos\frac{k}2\sum_{n=1}^{(S+1)/2}b_n n \sin^{2n-1}\frac{k}2 \notag \\
  =& \sin k\sum_{n=1}^{(S+1)/2}\frac{n}2 b_n \sin^{2n-2}\frac{k}2.
\end{align}
Next we express the density in Fourier space,
\begin{align}
    &n(x_l,x_m,x_n) \notag \\
    &= (\Delta \tilde{v})^3 \sum_{r,s,t} \underbrace{\left [ \frac{1}{N^3} \sum_{k_x, k_y, k_z}\tilde{f}(k_x,k_y,k_z, \tilde{v}_r, \tilde{v}_s, \tilde{v}_t)e^{i \frac{2 \pi}{N} l k_x} e^{i \frac{2 \pi}{N} m k_y} e^{i \frac{2 \pi}{N} n k_z}\right ]}_{= f(x_l, x_m, x_n, \tilde{v}_r, \tilde{v}_s, \tilde{v}_t)},
\end{align}
and apply three consecutive, exact, one-dimensional shifts,
\begin{align}
    \label{discreteXAdv}
     &n(x_l,x_m,x_n) = (\Delta \tilde{v})^3 \sum_{r,s,t} \frac{1}{N^3} \sum_{k_x, k_y, k_z} \tilde{f}(k_x, k_y, k_z, \tilde{v}_r, \tilde{v}_s, \tilde{v}_t) \notag \\
     &   (1 - i k_x d_1 + \tfrac{1}{2} k_x^2 d_1^2 + ...)e^{i \frac{2 \pi}{N} l k_x}(1 - i k_y d_2 + \tfrac{1}{2} k_y^2 d_2^2 + ...) e^{i \frac{2 \pi}{N} m k_y} \notag \\
     &(1 - i k_z d_3 + \tfrac{1}{2} k_z^2 d_3^2 + ...) e^{i \frac{2 \pi}{N} n k_z}.
\end{align}
Employing the definition of the shift vector (\ref{vShift}) and expressing velocity moments at the temporal midpoints of the x-advection ($t=T+\Delta t/2$), Eq. (\ref{discreteXAdv}) reproduces Eq. (\ref{nCh6}) in discretized form,
\begin{align}
    \label{nCh7}
    &\langle n_{s+1}(x_l, x_m, x_n) \rangle = \langle n_s(x_l, x_m, x_n) \rangle \notag \\
    &\qquad - \partial_x \langle (\bm{j}_s^*)_x(\tfrac{\Delta t}{2}) \rangle (x_l, x_m, x_n) c_1 \Delta t \notag \\
    &\qquad + \tfrac{1}{2} \partial_x^2 (\langle (\bm{\Pi}_{s-1}^\dagger)_{xx} \left (\tfrac{\Delta t}{2}\right )\rangle )(x_l, x_m, x_n) c_1^2 \Delta t^2 + O(\Delta t^3) .
\end{align}
Crucially, the BSL6D interpolator does not approximate the derivatives in Eq. (\ref{nCh7}) accurately. To eliminate these errors, BSL6D solves Eq. (\ref{nCh7}) in Fourier space by replacing the analytical operators $i k$ and $- k^2$ with their numerical counterparts $\alpha(k, S)$ and $\beta(k, S)$, respectively. The corrected expression for $\langle \phi \rangle$ is then obtained via an inverse Fourier transformation.

\subsubsection{Numerical illustration of the corrected scheme}

\begin{figure}[htbp] 
    \centering
    \begin{subfigure}[t]{0.75\columnwidth}
        \centering
        \includegraphics[width=\textwidth]{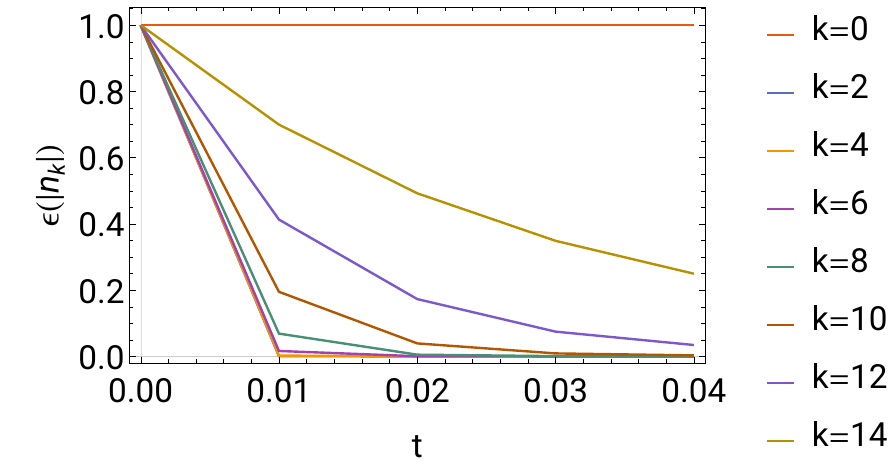}
        \caption{Relative error $\epsilon$ of the absolute value of the Fourier modes of the ion density $n$ without interpolation error correction.}
        \label{fig:damping_no_correction}
        \vspace{0.5cm}
    \end{subfigure}
    \begin{subfigure}[t]{0.75\columnwidth}
        \centering
        \includegraphics[width=\textwidth]{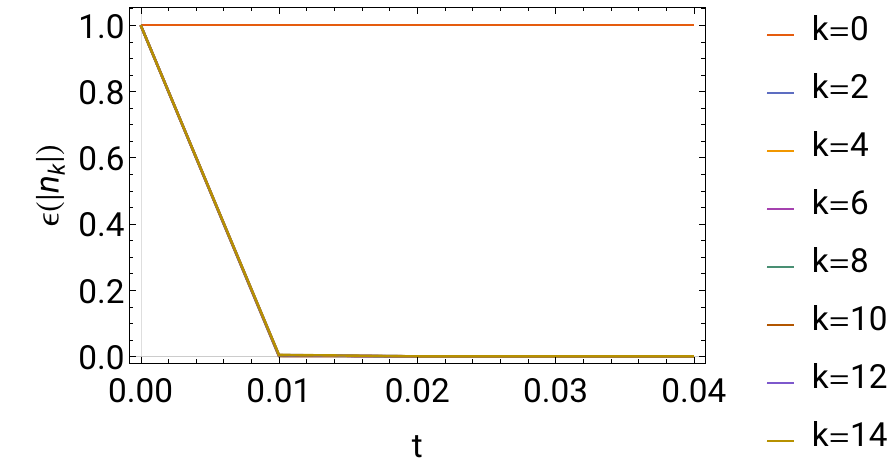}
        \caption{Relative error $\epsilon$ of the absolute value of the Fourier modes of the ion density $n$ with interpolation error correction.}
        \label{fig:damping_with_correction}
    \end{subfigure}
    \caption{Illustration of the interpolation error correction. The ion distribution function was initialized as $f = n_0 (1 + \alpha \delta n)f_{\text{M}}(v)$ with a constant electron background $\langle n_e \rangle = n_0$ on a 1D2V grid ($32 \times 65 \times 65$ points, domain $[0,2\pi] \times [-8,8]\times [-8,8]$). Here, $\delta n$ denotes a random density perturbation with magnitude $\alpha = 10^{-4}$. As there is only one spatial dimension in the x-direction, $\langle n \rangle = n$.}
    \label{fig:damped_scheme}
\end{figure}

To illustrate the interpolation error correction scheme, we first examine the damping of a random, non-quasi-neutral density perturbation. Without correction, the damping degrades significantly at high wavenumbers (Fig. \ref{fig:damping_no_correction}). 

\subsubsection{Long-time stability for large background density variations} 

\begin{figure}[!t] 
    \centering
    \begin{subfigure}[t]{0.75\columnwidth}
        \centering
        \includegraphics[width=\textwidth]{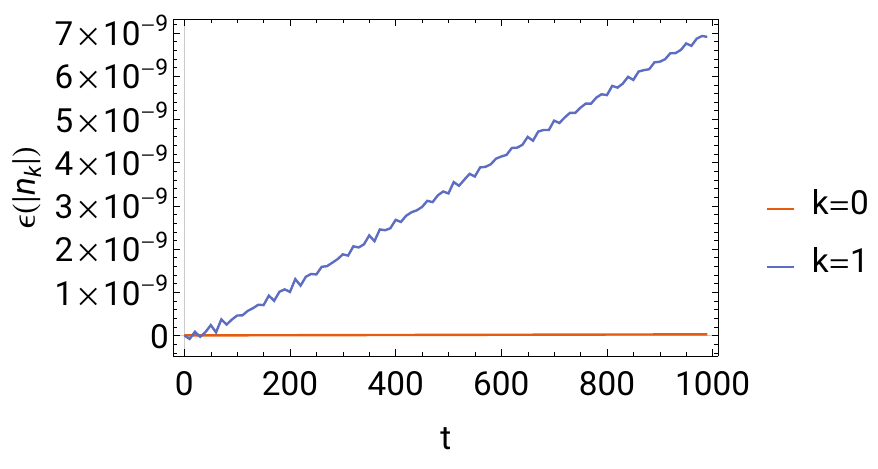}
        \caption{Relative error of the absolute value of the Fourier modes of the ion density $n$.}
        \label{fig:n_big_var_k_1}
        \vspace{0.5cm}
    \end{subfigure}
    \begin{subfigure}[t]{0.75\columnwidth}
        \centering
        \includegraphics[width=\textwidth]{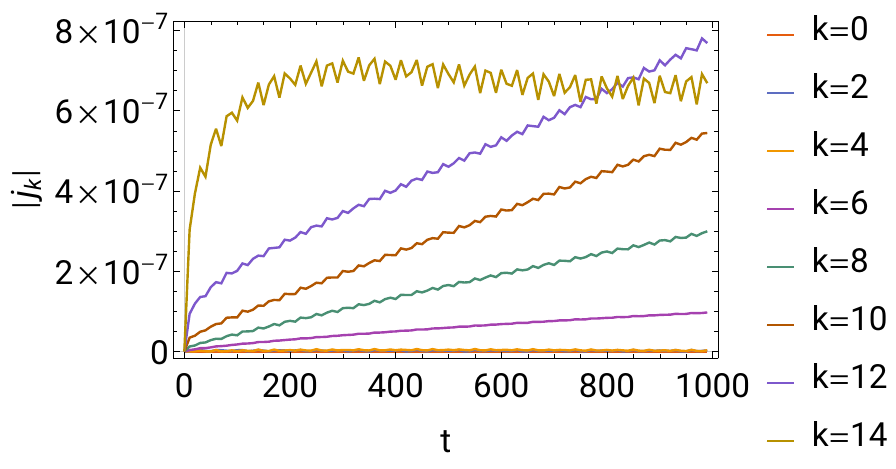}
        \caption{Absolute values of the Fourier modes of the current density for all wavenumbers up to the Nyquist frequency.}
        \label{fig:j_big_var_k_1}
    \end{subfigure}
    \caption{Relative error of the ion density and the absolute value of the Fourier modes of the current density. Here, $\delta n = \sin(x)$. Simulation parameters otherwise as in Fig. \ref{fig:damped_scheme}.}
    \label{fig:big_var_a}
\end{figure}

\begin{figure}[!t] 
    \centering
    \begin{subfigure}[t]{0.75\columnwidth}
        \centering
        \includegraphics[width=\textwidth]{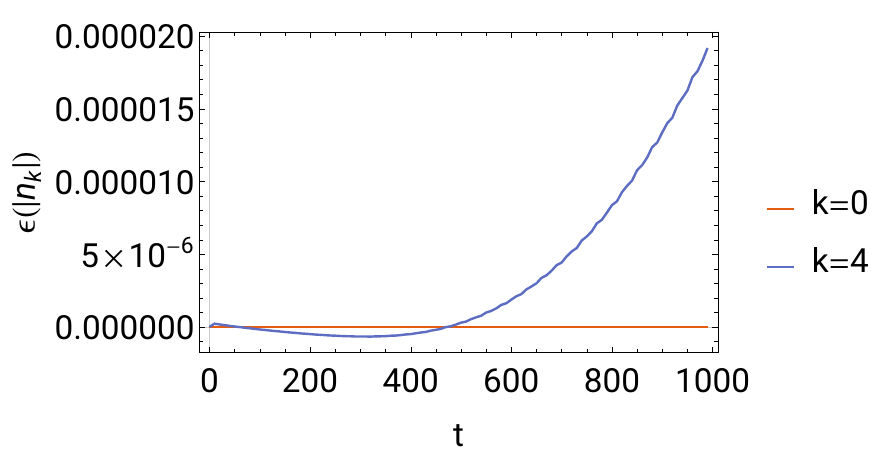}
        \caption{Relative error of the absolute value of the Fourier modes of the ion density $n$.}
        \label{fig:n_big_var_k_4}
         \vspace{0.5cm}
    \end{subfigure}
    \begin{subfigure}[t]{0.75\columnwidth}
        \centering
        \includegraphics[width=\textwidth]{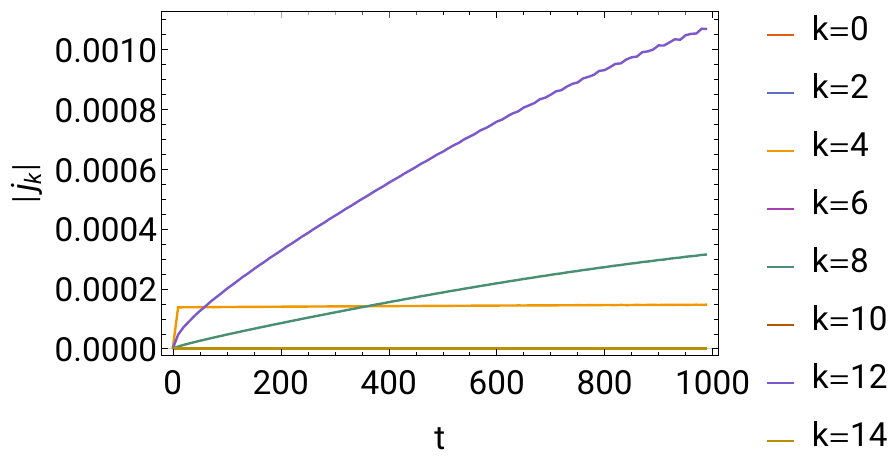}
        \caption{Absolute values of the Fourier modes of the current density for all wavenumbers up to the Nyquist frequency.}
        \label{fig:j_big_var_k_4}
    \end{subfigure}
    \caption{Relative error of the ion density and absolute Fourier modes of the current density for $\delta n = \sin(4x)$. Simulation parameters as in Fig. \ref{fig:damped_scheme}. The increased wavenumber leads to enhanced numerical errors.}
    \label{fig:big_var_b}
\end{figure}

To assess the stability of the corrected field solver under strong gradients, we initialize the system with large variations in the electron background. The ion density is set to $f = n_0 (1 + \alpha \delta n)f_{\text{M}}(v)$ and the electron background to $n_e = n_0(1 + \alpha \delta n)$, with modulation amplitude $\alpha = 0.95$ and $\delta n = \sin(x)$ (Fig. \ref{fig:big_var_a}) or $\delta n = \sin(4 x)$ (Fig. \ref{fig:big_var_b}), respectively. 

A fundamental property of Lagrange interpolation is the lack of commutation between the squared first derivative and the second-order derivative, i.e., $(\tilde{\partial}_x)^2 \neq \tilde{\partial_x^2}$. Consequently, the discrete continuity equation $\partial_t \langle n \rangle + \partial_x \langle j_x \rangle = 0$ cannot be satisfied identically if the density is constrained. In our setup, the fixed density necessitates an erroneous current density. As illustrated in Fig. \ref{fig:big_var_a}, this error introduces a weak numerical instability. For decreasing gradient length scale, the instability is exacerbated, as shown in Fig. \ref{fig:big_var_b}.
To ensure long-term stability and suppress these spurious effects, sufficient spatial resolution is required. Empirical analysis suggests that the absolute value of Fourier modes exceeding one-quarter of the Nyquist frequency should remain below the threshold $\kappa = 0.1$. This threshold was determined empirically by varying the grid resolution across several test cases.

\subsection{Convergence}
\label{sec:convergence}

Various theorems have been proven for the Vlasov-Poisson initial value problem \cite{glassey1996}, most notably the global existence and uniqueness of solutions in three dimensions for continuously differentiable, compactly supported initial data \cite{pfaffelmoser1992}. To the best of our knowledge, analogous results do not exist for the quasi-neutral two-species model considered here; consequently, we must impose certain regularity assumptions on the exact solution. First, we restrict the analysis to quasi-neutral initial conditions (i.e., $n_i = n_e$) whose time evolution satisfies
\begin{equation}
    \label{fSpace}
    f \in \mathcal{X} \coloneq \mathcal{C}^2_b(0,T; \mathcal{C}_{c, \: per_x}^{\infty}(\mathbb{R}^3_v \times \mathbb{R}^3_x)) .
\end{equation}
Here, $\mathcal{X}$ denotes the space of functions that are twice continuously differentiable and bounded on the time interval $[0,T]$, smooth and periodic in the spatial variables, and smooth and compactly supported in the velocity domain. The exact electric field $\bm{E}$, derived from the model introduced in Section \ref{sec:massless limit of drift-kinetic electrons}, is defined by
\begin{equation}
    \label{phiCFSAanalytic}
    \bm{E}:= -\langle \partial_x \phi \rangle \hat{x} - \frac{T_e}{e} \nabla (\ln(n) - \langle \ln (n) \rangle) ,
\end{equation}
where 
\begin{equation}
    \label{phiXCFSAanalytic}
     \langle \partial_x \phi \rangle := - \frac{1}{\langle n \rangle} \partial_x \langle \bm{\Pi}_{xx} \rangle - \frac{T_e}{e} \frac{1}{\langle n\rangle} \left \langle n \partial_x (\ln(n) - \langle \ln(n) \rangle )\right \rangle + \alpha \langle n_e - n \rangle.
\end{equation}
Except for the term proportional to $\alpha$, Eq. (\ref{phiXCFSAanalytic}) follows from the requirement $\partial_t \langle n \rangle = \partial_t^2 \langle n \rangle = 0$, which, when applied to the first moment of Eq. (\ref{vlasovEqRot}), yields
\begin{equation}
    0 = - \partial_x \langle \bm{\Pi}_{xx} \rangle + \langle E_x n \rangle .
\end{equation}
The term proportional to $\alpha$ is a damping parameter that penalizes ion density deviations from $\langle n_e \rangle$. Its precise value is irrelevant; we only require that the exact electric field restores ion density deviations from quasi-neutrality, i.e. non-quasi-neutral distribution functions with $\langle n_e \rangle \neq \langle n \rangle$ are attracted smoothly and injectively to a unique quasi-neutral distribution function in $\mathcal{X}$. More precisely, for an arbitrary, but fixed time $\tau \in [0,T]$ and a (possibly non-quasi-neutral) distribution function $g(\bm{x},\bm{v},\tau) \in \mathcal{C}_{c, \: per_x}^{\infty}(\mathbb{R}^3_v \times \mathbb{R}^3_x)$ satisfying 
\begin{equation}
    \label{globalErrorAssumptionA}
    \lVert f(\bm{x},\bm{v},\tau) - g(\bm{x},\bm{v},\tau) \rVert_\infty \leq \rho , \qquad \rho \in \mathbb{R}^+,
\end{equation}
we require that $g \in \mathcal{X}$ for $\tau < t < T$ and that a constant $L>0$ exists such that
\begin{equation}
    \label{globalErrorAssumptionB}
   \lVert f(\bm{x},\bm{v},t) - g(\bm{x},\bm{v},t) \rVert_\infty \leq \rho e^{L (t-\tau)}, \quad \text{for all} \quad 0 \leq \tau \leq t \leq T .
\end{equation}
Equations (\ref{globalErrorAssumptionA}) and (\ref{globalErrorAssumptionB}) guarantee the existence of a local stability region around the exact solution $f$, which is essential to prove the global convergence order of our method and is motivated by the standard stability analysis of ordinary differential equations, specifically Theorem 10.2 in \cite{hairer}. Intuitively, this regularization plays a role analogous to the Lipschitz continuity of the increment function in the stability theory of ordinary differential equations.  We remark that the exact electric potential $\phi$ (and thus $\bm{E}$) inherits its regularity from $f$, such that
\begin{equation}
\label{ESpace}
\phi \in \mathcal{C}^2_b(0,T; \mathcal{C}_{per_x}^{\infty}(\mathbb{R}^3_x)) .
\end{equation}
Similarly, the electric potential associated with the state $g$ inherits its regularity from $g$. Under these assumptions, we prove the existence of a constant $C$ such that 
\begin{equation}
    \label{globErrorBound}
    \lVert f(\bm{x}, \bm{\tilde{v}}, T) - f_s(\bm{x}, \bm{\tilde{v}}) \rVert_\infty \leq C \Delta t^2 ,
\end{equation}
where $f(\cdot, T)$ is the exact distribution function at time $t=T = s \Delta t$ and $f_s(\bm{x}, \bm{\tilde{v}})$ is the corresponding numerical solution. A similar convergence proof, incorporating interpolation effects, was established in \cite{besse2004} for the Vlasov-Poisson system.

\subsubsection{Initial step}
\label{sec:initial step}

We start with the initial step and prove that a constant $C_0$ exists such that
\begin{equation}
    \label{errorBoundInitStepA}
    \lVert f(\bm{x}, \bm{\tilde{v}}, \Delta t) - f_1(\bm{x}, \bm{\tilde{v}}) \rVert_\infty \leq C_0 \Delta t^2 .
\end{equation}
Let $(\bm{X}, \bm{V})$ denote the exact characteristics and $(\bm{\bar{X}},\bm{\bar{V}})$ denote the numerical characteristics. Furthermore, define $\bm{\xi} \coloneq (\bm{X}, \bm{V})$ and $\bm{\bar{\xi}} \coloneq (\bm{\bar{X}}, \bm{\bar{V}})$. It follows that 
\begin{align}
    \label{errorBoundInitStepB}
    &\lVert f(\bm{x}, \bm{\tilde{v}}, \Delta t) - f_1(\bm{x}, \bm{\tilde{v}}) \rVert_\infty \notag \\
    & \qquad = \lVert f(\bm{X}(\bm{x}, \bm{\tilde{v}}, - \Delta t), \bm{V}(\bm{x}, \bm{\tilde{v}}, - \Delta t), 0) \notag \\
    & \qquad \quad - f(\bm{\bar{X}}(\bm{x}, \bm{\tilde{v}}, - \Delta t), \bm{\bar{V}}(\bm{x}, \bm{\tilde{v}}, - \Delta t), 0) \rVert _\infty \notag \\
    & \qquad \leq L_0 \lVert \bm{\xi} - \bm{\bar{\xi}}\rVert_\infty , 
\end{align}
where the existence of the Lipschitz constant $L_0$ is guaranteed by the smoothness, spatial periodicity, and compact velocity support of the initial distribution $f$. To prove Eq. (\ref{errorBoundInitStepA}) using Eq. (\ref{errorBoundInitStepB}), we show that for any point $(\bm{x},\bm{\tilde{v}})$, there exist constants $C^0_X$ and $C^0_V$ such that
\begin{equation}
    \label{xCharBound}
    \lVert \bm{X}(\bm{x}, \bm{\tilde{v}}, - \Delta t) - \bm{\bar{X}}(\bm{x},\bm{\tilde{v}}, - \Delta t) \rVert_\infty \leq C^0_X \Delta t^3 
\end{equation}
and
\begin{equation}
    \label{vCharBound}
    \lVert \bm{V}(\bm{x}, \bm{\tilde{v}}, - \Delta t) - \bm{\bar{V}}(\bm{x},\bm{\tilde{v}}, - \Delta t) \rVert_\infty \leq C^0_V \Delta t^2 .
\end{equation}
Once these bounds are established, Eq. (\ref{errorBoundInitStepA}) follows with $C_0 = L_0 C^0_V$. We first prove Eq. (\ref{xCharBound}). Utilizing Eqs. (\ref{charactcharacteristicXRot}) and (\ref{charactcharacteristicVRot}), we derive the Taylor expansion of the exact spatial characteristic,
\begin{align}
    \label{xCharacteristicTaylor}
    &\bm{X}(\bm{x}, \bm{\tilde{v}}, - \Delta t) \notag \\
    &= \bm{x} + \int_0^{-\Delta t} dt \bm{R}(t) \left [ \bm{\tilde{v}} + \int_0^t dt' \bm{R}^{-1}(t') \bm{E}(0) \right ] + R_x(\Delta t) ,
\end{align}
where the Taylor remainder \cite{koenigsberger2004} is defined as
\begin{equation}
    \label{taylorRemainderX}
    R_x(\Delta t)  \coloneq \frac{1}{3!} \int_0^{- \Delta t}(\Delta t - \theta)^2 \frac{d^3}{d\theta^3} \bm{X}(\bm{x}, \bm{\tilde{v}}, T + \theta) d \theta.
\end{equation}
The third derivative,
\begin{equation}
    \label{taylorRemainderXexact}
    \frac{d^3}{dt^3} \bm{X} (\bm{x}, \bm{\tilde{v}},t) = \bm{R} \bm{\tilde{v}} \cdot \left [ - \omega_{ci}^2 + \nabla \bm{E} \right ] + (\partial_t \bm{R}) \bm{R}^{-1} \bm{E} + \partial_t \bm{E} ,
\end{equation}
is bounded due to the regularity of $f$; thus a constant $r_X$ exists such that $\lVert R_x(\Delta t) \rVert_\infty \leq r_X \Delta t^3$. Conversely, the numerical characteristic $\bm{\bar{X}}$ (see Section \ref{sec:bsl6d algorithm}) is given by
\begin{equation}
    \label{xCharA}
    \bm{\bar{X}}(\bm{x}, \bm{\tilde{v}}, - \Delta t) = \bm{x} + \int_0^{- \Delta t} dt \bm{R}(t) \bm{\bar{V}}(\bm{x}, \bm{\tilde{v}},- \Delta t),
\end{equation}
where 
\begin{equation}
    \label{xCharB}
    \bm{\bar{V}}(\bm{x}, \bm{\tilde{v}},- \Delta t) = \bm{\tilde{v}} + \int_0^{- \frac{\Delta t}{2}} dt \bm{R}^{-1}(t) \bm{E}_0(\bm{x}).
\end{equation}
Here, $\bm{E}_0$ is calculated from $f_0$ according to Eq. (\ref{phiCFSA}). Combining Eqs. (\ref{xCharA}) and (\ref{xCharB}) yields
\begin{align}
    \label{xInvCharNum}
    &\bm{\bar{X}} (\bm{x}, \bm{\tilde{v}}, - \Delta t) \notag \\
    &= \bm{x} + \int_0^{- \Delta t} dt \bm{R}(t) \left [ \bm{\tilde{v}} - \int_0^{- \frac{\Delta t}{2}} dt' \bm{R}^{-1}(t') \bm{E}_0(\bm{x}) \right ] ,
\end{align}
and comparing Eq. (\ref{xInvCharNum}) with Eq. (\ref{xCharacteristicTaylor}) leads to
\begin{align}
    &\lVert \bm{X}(\bm{x},\bm{\tilde{v}},-\Delta t) - \bm{\bar{X}}(\bm{x},\bm{\tilde{v}},-\Delta t) \rVert_\infty = \notag \\
    & \qquad = \left \lVert \int_0^{-\Delta t} dt \int_0^t dt' \bm{R}(t) \bm{R}(t') \bm{E}(0) + R_x(\Delta t) \right. \notag \\
    & \qquad \quad - \left. \int_0^{-\Delta t} dt \int_0^{-\frac{\Delta t}{2}} dt'\bm{R}(t) \bm{R}(t') \bm{E}_0 \right \rVert_\infty \notag \\
    & \qquad \leq \Delta t^2 \lVert \bm{E}(0) -\bm{E}_0 \rVert_\infty + \lVert R_x(\Delta t) \rVert_\infty .
\end{align}
To estimate $\lVert \bm{E}(0) -\bm{E}_0 \rVert_\infty$, we utilize the quasi-neutrality of the initial state, whereby Eq. (\ref{initPhiXCFSA}) implies 
\begin{equation}
    \label{E0num}
    \bm{E}_0 = - \langle \partial_x \phi_0 \rangle \hat{x} - \frac{T_e}{e} \nabla (\ln(n_0) - \langle \ln(n_0) \rangle) ,
\end{equation}
with 
\begin{align}
    -\langle \partial_x \phi_0 \rangle =& \frac{1}{\langle n_0 \rangle}\partial_x \left \langle (\bm{\Pi}_0)_{xx} \left(\tfrac{\Delta t}{2}\right)\right \rangle \notag \\
    &+ \frac{T_e}{e \langle n_0 \rangle} \langle n_0 \mathcal{D}_0 (\ln(n_0) - \langle \ln(n_0) \rangle) \rangle .
\end{align}
Comparing $\bm{E}_0$ (\ref{E0num}) to Eq. (\ref{phiXCFSAanalytic}) with $\alpha = 0$, and invoking the regularity of both electric fields, we find that
\begin{equation}
    \label{errorE0}
    \lVert \bm{E}(0) - \bm{E}_0 \rVert_\infty \leq C^0_E \Delta t
\end{equation}
since
\begin{equation}    
    \lVert \partial_x \left \langle \bm{\Pi}_{xx} \left ( \tfrac{\Delta t}{2}\right ) \right \rangle - \partial_x \left \langle \bm{\Pi}_{xx} \left (0 \right ) \right \rangle \rVert = O(\Delta t),
\end{equation}
which confirms Eq. (\ref{xCharBound}). We now prove Eq. (\ref{vCharBound}). Using Eqs. (\ref{charactcharacteristicXRot}) and (\ref{charactcharacteristicVRot}), we expand the analytical velocity characteristic, 
\begin{align}
    \label{vCharacteristicTaylor}
    &\bm{V}(\bm{x}, \bm{\tilde{v}}, - \Delta t) = \notag \\
    &\bm{\tilde{v}} + \int_0^{- \Delta t} dt \bm{R}^{-1}(t) \left [\bm{E}(0) + \int_0^t dt' \left (\partial_{t'} \bm{E}(0) + \bm{R}(t') \bm{\tilde{v}} \cdot \nabla \bm{E} (0)\right ) \right ] \notag \\
    & + R_v(\Delta t),
\end{align}
where $R_v(\Delta t)$ is defined analogous to (\ref{taylorRemainderX}) by replacing $\bm{x} \rightarrow \bm{v}$. The third derivative,
\begin{align}
    \frac{d^3}{dt^3} \bm{V} (\bm{x}, \bm{\tilde{v}},t) =& \: \bm{R}^{-1} \left ( \partial_t^2 \bm{E} - \omega_{ci}^2 \bm{E} \right ) + \partial_t \bm{R}^{-1} \left ( \bm{v} \cdot \nabla \bm{E} + 2 \partial_t \bm{E} \right ) \notag \\
    &+ \bm{v} \cdot \nabla \left ( \bm{v} \cdot \nabla \bm{E} \right ) + \left (\partial_t \bm{R} \bm{\tilde{v}} + \bm{E} \right ) \cdot \nabla \bm{E} ,
\end{align}
is bounded by the regularity of $f$, ensuring a constant $r_V$ exists such that $\lVert R_v(\Delta t) \rVert_\infty \leq r_V \Delta t^3$. For the numerical characteristic $\bm{\bar{V}}$ obtained from the Strang-splitting (see Section \ref{sec:bsl6d algorithm}), we have
\begin{equation}
    \bm{\bar{V}}(\bm{x},\bm{\tilde{v}},- \Delta t) = \bm{\tilde{v}}_\alpha + \int_{- \frac{\Delta t}{2}}^{-\Delta t} dt \bm{R}^{-1}(t) \bm{E}_0(\bm{\bar{X}}(\bm{x}, \bm{\tilde{v}}_\alpha, - \Delta t)) ,
\end{equation}
where $\bm{\tilde{v}}_\alpha$ is the velocity after the initial $-\Delta t/2$ step, and $\bm{\bar{X}}(\bm{x}, \bm{\tilde{v}}_\alpha, -\Delta t)$ is the position after the $-\Delta t$ spatial step. It follows from (\ref{charactcharacteristicXRot}) and (\ref{charactcharacteristicVRot}) that
\begin{equation}
    \bm{\tilde{v}}_\alpha = \bm{\tilde{v}} + \int_0^{-\frac{\Delta t}{2}} dt \bm{R}^{-1}(t) \bm{E}_1(\bm{x}) 
\end{equation}
and
\begin{equation}
    \bm{\bar{X}}(\bm{x}, \bm{\tilde{v}}_\alpha, - \Delta t) = \bm{x} + \int_0^{-\Delta t} dt \bm{R}(t) \bm{\tilde{v}}_\alpha .
\end{equation}
Thus,
\begin{align}
    \label{vInvCharNum}
    &\bm{\bar{V}}(\bm{x},\bm{\tilde{v}}, - \Delta t) \notag \\
    &= \bm{\tilde{v}} + \int_0^{-\frac{\Delta t}{2}} dt \bm{R}^{-1}(t) \bm{E}_1(\bm{x}) \notag \\
    &\quad + \int_{-\frac{\Delta t}{2}}^{-\Delta t} dt \bm{R}^{-1}(t) \left [ \bm{E}_0(\bm{x}) + \int_0^{-\Delta t} dt' \bm{R}(t') \bm{\tilde{v}}_\alpha \cdot \nabla \bm{E}_0(\bm{x}) \right ] \notag \\
    &= \bm{v} + \int_0^{- \Delta t} dt \bm{R}^{-1}(t) \bm{E}_0(\bm{x}) \notag \\
    & \quad + \int_{- \frac{\Delta t}{2}}^{- \Delta t} dt \int_0^{- \Delta t} dt'  \bm{R}(t') \bm{\tilde{v}}_\alpha \cdot \nabla \bm{E}_0 (\bm{x}) \notag \\
    & \quad + \int_0^{- \Delta t} dt \bm{R}^{-1}(t) ( \bm{E}_1 - \bm{E}_0 ).
\end{align}
Since $f_1$ is obtained via smooth, bounded shifts from $f_0$, we have $f_1 \in \mathcal{C}_{c, \: per_x}^\infty(\mathbb{R}^3_x \times \mathbb{R}_v^3)$, implying $\bm{E}_1 \in \mathcal{C}_{per_x}^\infty(\mathbb{R}^3_x)$. Due to this regularity, there exists a constant $\bar{C}_E^0$ such that 
\begin{equation}
    \label{errorE1}
    \lVert \bm{E}_1 - \bm{E}_0 \rVert \leq \bar{C}_E^0 \Delta t .
\end{equation}
Comparing (\ref{vInvCharNum}) with (\ref{vCharacteristicTaylor}) and invoking the regularity of all quantities, we establish the bound (\ref{vCharBound}) 
\begin{equation}
    \lVert \bm{V}(\bm{x}, \bm{\tilde{v}}, - \Delta t) - \bm{\bar{V}}(\bm{x},\bm{\tilde{v}},- \Delta t) \rVert_\infty \leq C^0_V \Delta t^2 . 
\end{equation}
This proves Eq. (\ref{errorBoundInitStepA}). However, the estimate (\ref{errorBoundInitStepA}) alone is insufficient to establish the global second-order error convergence due to error accumulation over multiple time steps. The key feature of our field solver is that it achieves enhanced accuracy for the electric field once it adapts to a slightly perturbed state of the distribution function. This improved accuracy is crucial for recovering the global second-order convergence.

\subsubsection{Induction}
\label{sec:induction}

We now show that for all steps $s>0$, the accuracy of the electric field (\ref{phiCFSA}) exceeds the previous estimates (\ref{errorE0}) and (\ref{errorE1}). Specifically, there exist constants $C_E^s$ such that
\begin{equation}
    \label{errorEs}
    \lVert \bm{E}(T) - \bm{E}_s \rVert_\infty \leq C_E^s \Delta t^2 .
\end{equation}
We start with $s=1$. The difficulty of the proof lies in the fact that we determine the electric potential in the middle of a Strang-splitting step; thus, care must be taken when assessing the consistency order of the different velocity moments occurring in Eq. (\ref{phiXCFSA}). Since the density does not change during the final v-advection of the Strang splitting, it is already in its final state when the electric field is computed (after the x-advection). For the adiabatic part of the potential, it follows that a constant $C^1_{E - \langle E \rangle}$ exists such that
\begin{align}
    \label{EtildeBound}
    \lVert \bm{E}_1 - \langle \bm{E}_1 \rangle \rVert_\infty &= \left \lVert - \frac{T_e}{e} \nabla (\ln(n_1) - \langle \ln(n_1) \rangle ) \right \rVert_\infty \notag \\
    &\leq C^1_{E - \langle E \rangle} \Delta t^3 ,
\end{align}
as the accuracy of the fluctuating density (i.e., the density with zero flux-surface average) is only constrained by the Strang-splitting accuracy in the first step. Note that the estimate (\ref{EtildeBound}) will not improve further. A similar argument cannot be made for the flux-surface averaged component of the electric field, and we therefore employ an indirect approach. Recall from Eq. (\ref{phiXCFSA}) that
\begin{align}
    \label{phiIntA}
    &- \langle \partial_x \phi_1 \rangle \langle n_1 \rangle c_2 \Delta t  = \notag \\
    & \quad - \left \langle (\bm{j}_0^\dagger)_x \left (\tfrac{3}{2} \Delta t \right ) \right \rangle + \frac{c_1}{2} \partial_x \left \langle (\bm{\Pi}_0^\dagger)_{xx} \left (\tfrac{3}{2} \Delta t \right ) \right \rangle \notag \\
    & \quad + \frac{1}{c_1 \Delta t} \int dx \left \langle n_1 - n_e \right \rangle + \frac{T_e}{e} c_2 \Delta t \langle n_1 \mathcal{D} (\ln(n_1) - \langle \ln(n_1) \rangle) \rangle ,
\end{align}
where $\mathcal{D}$ was defined as 
\begin{equation*}
    \mathcal{D}= \left ( \partial_x  +  c_3 \partial_y \right ), \qquad \text{with} \qquad c_3 = \tan \left (\frac{\Omega \Delta t}{2} \right ).
\end{equation*}
Inserting (\ref{phiIntA}) into equation (\ref{jCh5}), which for $s=1$ takes the form
\begin{align}
    \left \langle (\bm{j}_1^*)_x \left (\tfrac{3}{2} \Delta t \right ) \right \rangle =& \left \langle (\bm{j}_0^\dagger )_x \left (\tfrac{3}{2} \Delta t \right ) \right \rangle - \langle \partial_x  \phi_1 \rangle \langle n_1 \rangle c_2 \Delta t \notag \\
    &- \frac{T_e}{e} c_2 \Delta t \left \langle n_1 \mathcal{D} \left (\ln(n_1) - \langle \ln(n_1) \rangle \right ) \right \rangle ,  
\end{align}
yields
\begin{equation}
    \label{phiIntB}
    \left \langle (\bm{j}_1^*)_x \left (\tfrac{3}{2} \Delta t \right ) \right \rangle = \frac{c_1}{2} \partial_x \left \langle (\bm{\Pi}_0^\dagger)_{xx} \left (\tfrac{3}{2} \Delta t \right ) \right \rangle + \frac{1}{c_1 \Delta t} \int dx \left \langle n_1 - n_e \right \rangle .
\end{equation}
Together with Eq. (\ref{nCh6}), which for $s=1$ reads
\begin{align}
    \langle n_2 \rangle - \langle n_1 \rangle =&  - \partial_x \left \langle (\bm{j}_1^*)_x \left (\tfrac{3}{2} \Delta t \right ) \right \rangle c_1 \Delta t \notag \\
    &+ \tfrac{1}{2}\partial^2_x \left \langle (\bm{\Pi}_0^\dagger)_{xx} \left (\tfrac{3}{2} \Delta t \right ) \right \rangle c_1^2 \Delta t^2 + O(\Delta t^3),
\end{align}
Eq. (\ref{phiIntB}) implies that 
\begin{equation}
    \label{nHelp}
    \langle n_2 \rangle - \langle n_e \rangle = O(\Delta t^3) .
\end{equation}
The $O(\Delta t^3)$ term originates from the influence of higher order velocity moments on the density, which we have so far neglected by truncating the density expansion at $O(\Delta t^3)$, see Eq. (\ref{nCh6}). We can therefore write Eq. (\ref{nHelp}) as
\begin{equation}
    \langle n_2 \rangle - \langle n_e \rangle = \langle r(\bm{M}) \rangle ,
\end{equation}
where $r(\bm{M})$ is a remainder function that summarizes the influence of these higher moments. Since $f_1$ is obtained via smooth, bounded shifts from $f_0$, it follows that $\lVert r(\bm{M}) \rVert_\infty$ is also bounded. As the variation of $\bm{M}$ between subsequent steps is proportional to $\Delta t$, we conclude that a constant $C_{\langle n \rangle}^1$ exists such that
\begin{equation}
    \label{deltaN}
    \langle n_2 \rangle - \langle n_1 \rangle = O(\Delta t^4), \quad \text{and} \quad \lVert n_2 - n_1 \rVert_\infty \leq C_{\langle n \rangle}^1 \Delta t^4.
\end{equation}
Ideally, the electric potential should fix $\langle n \rangle$ exactly (i.e., $\partial_t \langle n \rangle = 0$). However, as we have just shown, the definition of the electric potential (\ref{phiCFSA}) allows for an $O(\Delta t^4)$ error in the density change between the first and the second step (\ref{deltaN}). Since the electric potential can only influence the density via an $\Delta t^2$ factor, see Eq. (\ref{nCh6}) and (\ref{jCh4}), it follows that a constant $C^1_E$ exists such that
\begin{equation}
    \label{accuracyE0}
    \lVert \langle \bm{E}(\Delta t) \rangle - \langle \bm{E}_1 \rangle \rVert_\infty \leq C^1_E \Delta t^2 .
\end{equation}
In particular, Eq. (\ref{accuracyE0}), together with the regularity of $f_1$, implies that a constant $C^1_{\dot{E}}$ exists such that
\begin{equation}
    \label{ddtEBound}
    \lVert \langle \partial_t \bm{E}_1(\Delta t) \rangle - \langle (\partial_t\bm{E})_1 \rangle \rVert_\infty \leq C^1_{\dot{E}} , 
\end{equation}
where 
\begin{equation}
    \lVert \langle \partial_t\bm{E}_1 \rangle \rVert_\infty \coloneq \frac{1}{\Delta t} \lVert \langle \bm{E}_1 - \bm{E}_0 \rangle \rVert_\infty
\end{equation}
is the discretized time derivative of the flux-surface averaged electric field. Due to Eq. (\ref{ddtEBound}), the estimate (\ref{vCharBound}) cannot be improved for the $\bm{\bar{V}}$ characteristic of the second step since
\begin{flalign}
    \label{vInvCharNum1}
    &\bm{\bar{V}}(\bm{x},\bm{\tilde{v}}, - \Delta t) \notag \\
    &= \bm{\tilde{v}} + \int_0^{-\frac{\Delta t}{2}} dt \bm{R}^{-1}(t) \bm{E}_1(\bm{x}) \notag \\
    &\quad + \int_{-\frac{\Delta t}{2}}^{-\Delta t} dt \bm{R}^{-1}(t) \left [ \bm{E}_0(\bm{x}) + \int_0^{-\Delta t} dt' \bm{R}(t') \bm{\tilde{v}}_\alpha \cdot \nabla \bm{E}_0(\bm{x}) \right ] &\notag \\
    &= \bm{v} + \int_0^{- \Delta t} dt \bm{R}^{-1}(t) \bm{E}_0(\bm{x}) + \Delta t \int_{-\frac{\Delta t}{2}}^{\Delta t} dt (\partial_t \bm{E})_1 \notag \\
    &\quad + \int_{- \frac{\Delta t}{2}}^{- \Delta t} dt \int_0^{- \Delta t} dt'  \bm{R}(t') \bm{\tilde{v}}_\alpha \cdot \nabla \bm{E}_1 (\bm{x}) . &
\end{flalign}
Thus, our scheme collects another $O(\Delta t^2)$ error in the second step. However, the regularity of $f_1$ is preserved; i.e., $f_2 \in \mathcal{C}^2_b(0,T; \mathcal{C}_{c, \: per_x}^{\infty}(\mathbb{R}^3_v \times \mathbb{R}^3_x))$ and $\bm{E}_2 \in \mathcal{C}^2_b(0,T; \mathcal{C}_{per_x}^{\infty}(\mathbb{R}^3_x))$. Most importantly, for the subsequent step ($s=2$), the estimate (\ref{ddtEBound}) improves,
\begin{equation}
    \label{ddtE2Bound}
    \lVert \langle \partial_t \bm{E}_2(\Delta t) \rangle - \langle (\partial_t\bm{E})_2 \rangle \rVert_\infty \leq C^2_{\dot{E}} \Delta t ,
\end{equation}
and consequently 
\begin{equation}
    \label{vCharBoundS}
    \lVert \bm{V}(\bm{x}, \bm{\tilde{v}}, T - \Delta t) - \bm{\bar{V}}(\bm{x},\bm{\tilde{v}}, T - \Delta t) \rVert_\infty \leq C_V^3 \Delta t^3 .
\end{equation}
The order of the bounds (\ref{vCharBoundS}) and (\ref{xCharBound}) remains unchanged for all $s>2$, that is, our scheme requires two discrete steps to reach its numerical equilibrium after initialization. This follows from the fact that for all $s>2$, the density error $\langle n_{s+1}\rangle - \langle n_s \rangle$ remains $O(\Delta t^3)$ by the same argument as above, and consequently Eq. (\ref{ddtE2Bound}) holds for all $s>2$. To conclude the induction, it remains to be shown that the bounding constants $C_X^s$ and $C_V^s$ remain finite for finite simulation times. Since analytical results regarding stability, existence and uniqueness of solutions for our specific model are absent, we apply the technique described in Section II.3 of \cite{hairer}. The core idea is not to calculate the error propagation of local errors $\lVert e_s \rVert$ through the numerical scheme itself, but along stable analytic solutions in the vicinity of the exact solution to the final time $T$. This requires the existence of a local stability region around the exact solution for a given initial condition, which we have already stipulated in the introduction.

The difference from the treatment in  Section II.3 of \cite{hairer} lies in the fact that the numerical distribution function of our method consistently violates quasi-neutrality due to the inherent $O(\Delta t^3)$ density error after each step. However, as demonstrated, this violation is necessary to obtain the increased accuracy of the electric potential for $s>2$. For the induction, we identify an analytic, non-quasi-neutral state that aligns with the numerical solution at each step and propagate the error along this auxiliary solution. By assumption, this solution is guaranteed to exist and reside in $\mathcal{X}$. Consequently, the constants $C_X^s$ and $C_V^s$ exist in every step and are bounded. Next we define the local error 
\begin{equation}
    \label{locErrorBound}
    \lVert e_s \rVert \coloneq \lVert f(\bm{x}, \bm{\tilde{v}}, s\Delta t) - f_s(\bm{x}, \bm{\tilde{v}}) \rVert_\infty , \qquad \lVert e_s \rVert \leq C_s \Delta t^3,
\end{equation}
from which, using Eq. (\ref{globalErrorAssumptionB}), the corresponding error $E_s$ at the final time $T$ can be bounded,
\begin{equation}
    \lVert E_s \rVert_\infty \leq e^{L \Delta t(N-s)} \lVert e_s \rVert_\infty.
\end{equation}
Following \cite{hairer}, the global error is then given by
\begin{equation}
    \label{globalError}
    \lVert E \rVert_\infty \leq \sum_{i=1}^N \lVert E_i \rVert_\infty \leq \sum_{i=1}^N e^{L \Delta t(N-i)} \lVert e_i \rVert_\infty \leq \sum_{i=1}^Ne^{L \Delta t(N-i)} C_i \Delta t^3 ,
\end{equation}
and employing the estimate of the sum in Eq. (\ref{globalError}) given in Section II.3 of \cite{hairer}, we conclude that
\begin{equation}
    \lVert E \rVert_\infty \leq \sum_{i=1}^Ne^{L \Delta t(N-i)}\max_{i \in \{1,...,N\}}(C_i) \Delta t^3 \leq C \Delta t^2 .
\end{equation}
This establishes Eq. (\ref{globErrorBound}).

\paragraph*{Remark.}

Can $\langle \bm{E}_s \rangle$ be made more accurate by including the exact form of $\bm{M}$ at $O(\Delta t^3)$ or higher in the definition (\ref{phiXCFSA})? To address this, note that Eq. (\ref{deltaN}) implies that the electric potential eliminates the $O(\Delta t^3)$ density error (\ref{nCh6}) from the previous step. At the same time, the previous density error is recreated with possible $O(\Delta t^4)$ deviations. The elimination of an $O(\Delta t^3)$ density error up to $O(\Delta t^4)$ deviations requires a current density to be determined accurately up to $O(\Delta t^2)$, see Eq. (\ref{nCh2}). By including higher orders of $\bm{M}$ in the Eq. (\ref{phiXCFSA}), the error in (\ref{deltaN}) can indeed be made arbitrarily small. However, the resulting potential can never be physically more accurate than $O(\Delta t^2)$, because the physical accuracy of the current density is limited by the local $O(\Delta t^3)$ error of the Strang splitting. Put differently, a more complex electric potential would satisfy the quasi-neutrality condition more accurately, but it would do so in an unphysical manner. For the same reason, the adiabatic part of the electric potential (\ref{phiCFSA}) is likewise limited to $O(\Delta t^2)$ accuracy. 

\section{Numerical verification}
\label{sec:numerical verification}

\subsection{Error convergence of the electric potential} 
\label{sec:error convergence of the electric potential}

The verification of global second-order convergence for the proposed scheme relies on the second-order convergence of the electric potential. To numerically validate this, we employ a 1D1V unmagnetized configuration where the spatial domain is aligned with the $\hat{x}$ direction (orthogonal to the slab flux-surfaces). This setup specifically isolates the new, quasi-neutral coupling of kinetic ions to a static electron background. 

\begin{figure}[!t] 
    \centering
    \begin{subfigure}[t]{0.75\columnwidth}
        \centering
        \includegraphics[width=\textwidth]{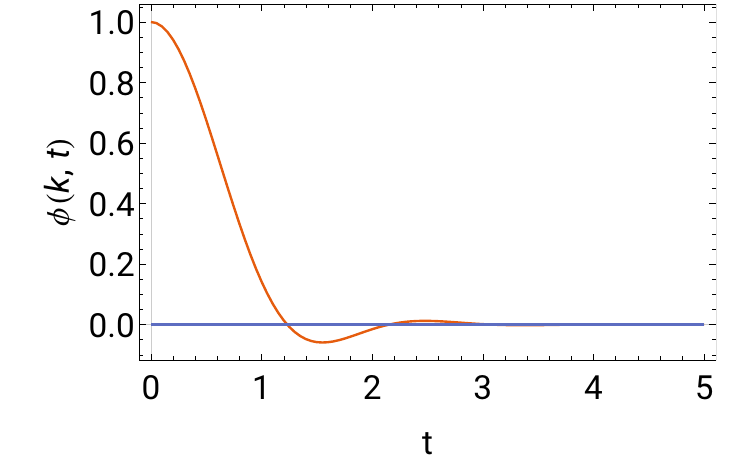}
        \caption{Real part (red) and imaginary part (blue) of the Fourier amplitude of the electric potential for an initial temperature perturbation profile $g(x) = \sin(x)$.}
        \label{fig:phi_sound_wave}
        \vspace{0.5cm}
    \end{subfigure}
    \begin{subfigure}[t]{0.75\columnwidth}
        \centering
        \includegraphics[width=\textwidth]{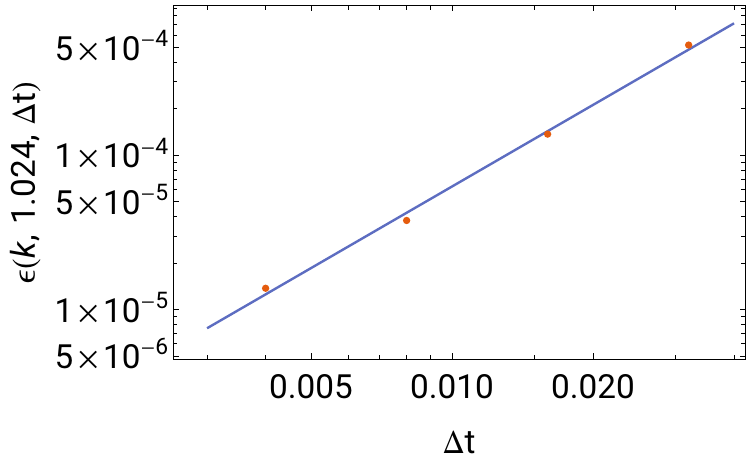}
        \caption{Relative error (\ref{relErrPhi}) for different timesteps, 
        with fitted scaling $\Delta t^{1.8}$}
    \label{fig:p_err_convergence}
    \end{subfigure}
    \caption{Illustration of the time evolution of the quasi-neutral electric potential for an initial temperature perturbation and its relative error. The fitted scaling of $\Delta t^{1.8}$ is consistent with the theoretical
    $O(\Delta t^2)$ prediction; the slight deviation is attributed to residual interpolation errors at the chosen evaluation time t=1.024.}
    \label{fig:electric_potential_error}
\end{figure}

Following the methodology described in \cite{pelkner}, we first construct a manufactured solution for this problem. Since any pure initial density perturbation would be immediately neutralized by the field-solver, we initialize a small temperature perturbation,
\begin{equation}
    \label{tempPert}
    f(x, v, t = 0) = \frac{n_0}{\sqrt{2 \pi} \sigma} e^{-\frac{v^2}{2 \sigma^2}}, \qquad \sigma \coloneq v_{\text{th}} \sqrt{1 + \alpha g(x)},
\end{equation}
where $\alpha = 10^{-4}$ denotes the dimensionless magnitude of the temperature perturbation, $g(x)$ is the spatial perturbation profile, and $v_{\text{th}}^2=T/m$ is the ion thermal velocity. Since $\alpha \ll 1$, the initial condition (\ref{tempPert}) can be linearized,
\begin{equation}
    f(x,v,t=0) = n_0 f_{\text{M}} + \alpha f_1 (x,v,t=0) + O(\alpha^2),
\end{equation}
where 
\begin{equation}
    \label{tempPertInitCond}
     f_1(x,v,t=0) = \frac{n_0}{2} f_{\text{M}}(v) \left(-1 + \frac{v^2}{v_{\text{th}}^2} \right) g(x) 
\end{equation}
and
\begin{equation}
    \label{maxwellian}
    f_{\text{M}}(v) \coloneq \frac{1}{\sqrt{2 \pi} v_{\text{th}}} e^{-\frac{v^2}{2 v_{\text{th}}^2}}.
\end{equation}
The time evolution of $f_1$ is governed by the linearized Vlasov equation \cite{goldston},
\begin{equation}
    \label{linearVlasov}
    \partial_t f_1 + v \partial_x f_1 - \frac{q n_0}{m} \partial_x \phi \partial_v f_{\text{M}} = 0 .
\end{equation}
For simplicity, we assume a planewave ansatz $g = e^{i k x}$, which reduces the spatial derivative to a multiplication by $i k$. To compute the causal response ($t>0$) of $f_1$ to the initial perturbation (\ref{tempPert}), we transform Eq. (\ref{linearVlasov}) to the frequency domain (for $\text{Im}(\omega)>0$) using
\begin{equation}
    \label{FT}
    \hat{h} (\omega) \coloneq \frac{1}{\sqrt{2 \pi}} \int_{-\infty}^{\infty} dt h(t) e^{-i \omega t},
\end{equation}
where $h$ and $\hat{h}$ indicate an arbitrary function and its respective Fourier transform. This gives
\begin{equation}
    \label{linearVlasovFou}
    -i(\omega - k v) \hat{f}_1 - \frac{q n_0}{m} i k \hat{\phi} \partial_v f_{\text{M}} = \frac{1}{\sqrt{2 \pi}} f_1(x,v,t=0) .
\end{equation}
By algebraically solving Eq. (\ref{linearVlasovFou}) for $\hat{f}_1$ and integrating over velocity space, we obtain an analytical expression for the linearized ion density perturbation $\hat{n}_1$. Invoking the quasi-neutrality constraint ($\hat{n}_1 = 0$) allows us to solve for the electric potential,
\begin{equation}
    \label{pLinFou}
    \hat{\phi}(k, \omega) = - \frac{i \alpha}{4 k \sqrt{\pi}} \frac{2 \xi - (1 - 2 \xi^2) Z (\xi)}{1 + \xi Z(\xi)}, \qquad \xi = \frac{\omega}{\sqrt{2} v_{\text{th}} k} ,
\end{equation}
where $Z$ is the plasma dispersion function \cite{nrl}. Following \cite{pelkner}, we construct the time domain solution for the electric potential from Eq. (\ref{pLinFou}), shown in Fig. \ref{fig:phi_sound_wave}. Based on this manufactured solution, the expected $O(\Delta t^2)$ error convergence of the numerical electric potential is recovered, as illustrated in Fig. \ref{fig:p_err_convergence} for $g(x) = \sin(x)$. Here, we have evaluated the relative error 
\begin{equation}
    \label{relErrPhi}
    \epsilon(k,t,\Delta t) \coloneq \left | 1 -\frac{\phi^{\text{num}}(k,t, \Delta t)}{\phi(k,t)} \right |, \qquad \text{for} \qquad k=1, \:t = 1.024,
\end{equation}
where $\phi^{\text{num}}$ is the numerical solution. The evaluation time  $t=1.024$ was chosen because it coincides with a discrete time step for all investigated values of $\Delta t$.

\subsection{Bernstein waves} 
\label{sec:bernstein waves}

\begin{figure}[!t] 
    \centering
    \begin{subfigure}[t]{0.75\columnwidth}
        \centering
        \includegraphics[width=\textwidth]{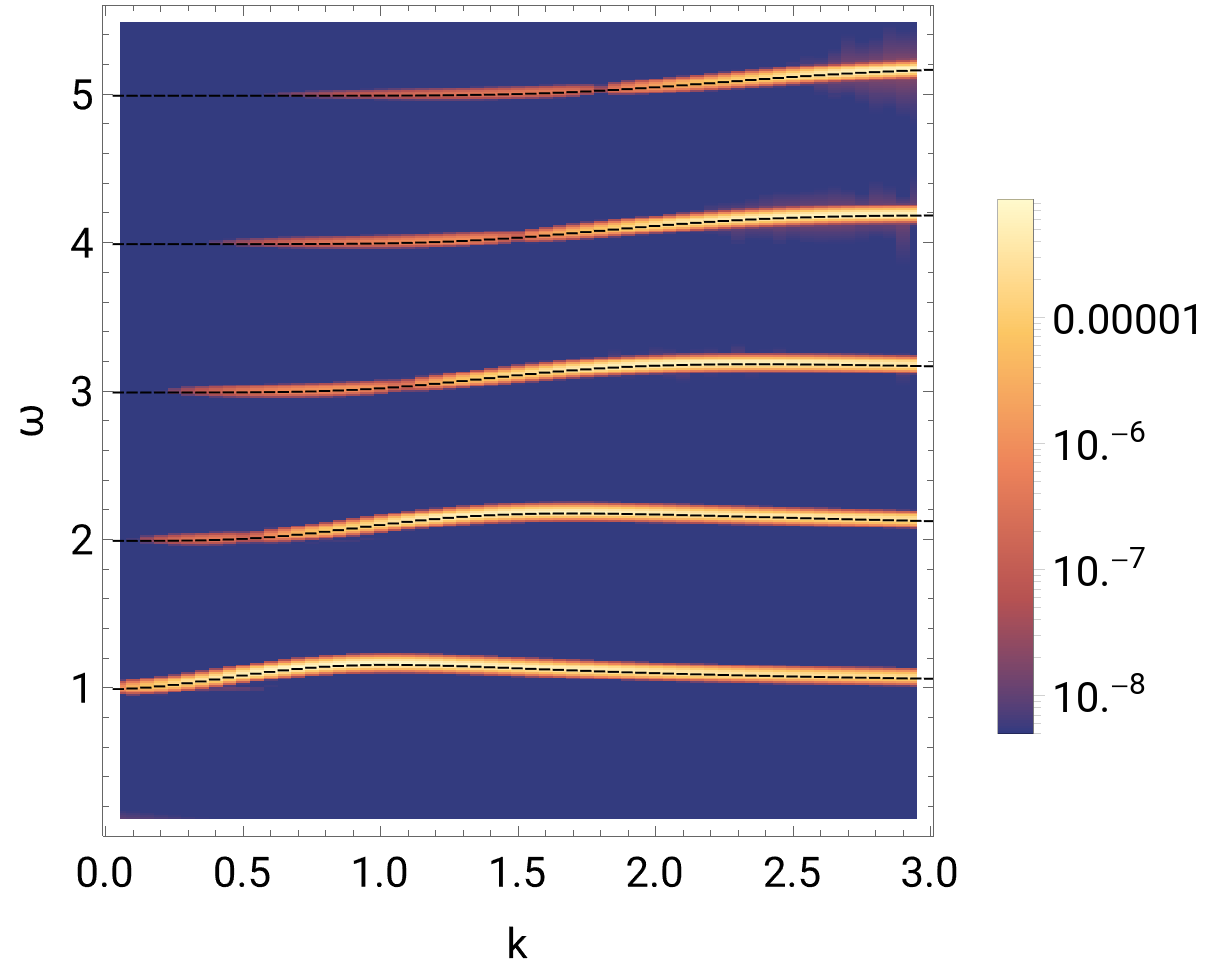}
        \caption{Neutralizing IBW ($k_y\neq 0$).}
        \label{fig:neut_ibw}
        \vspace{0.5cm}
    \end{subfigure}
    \begin{subfigure}[t]{0.75\columnwidth}
        \centering
        \includegraphics[width=\textwidth]{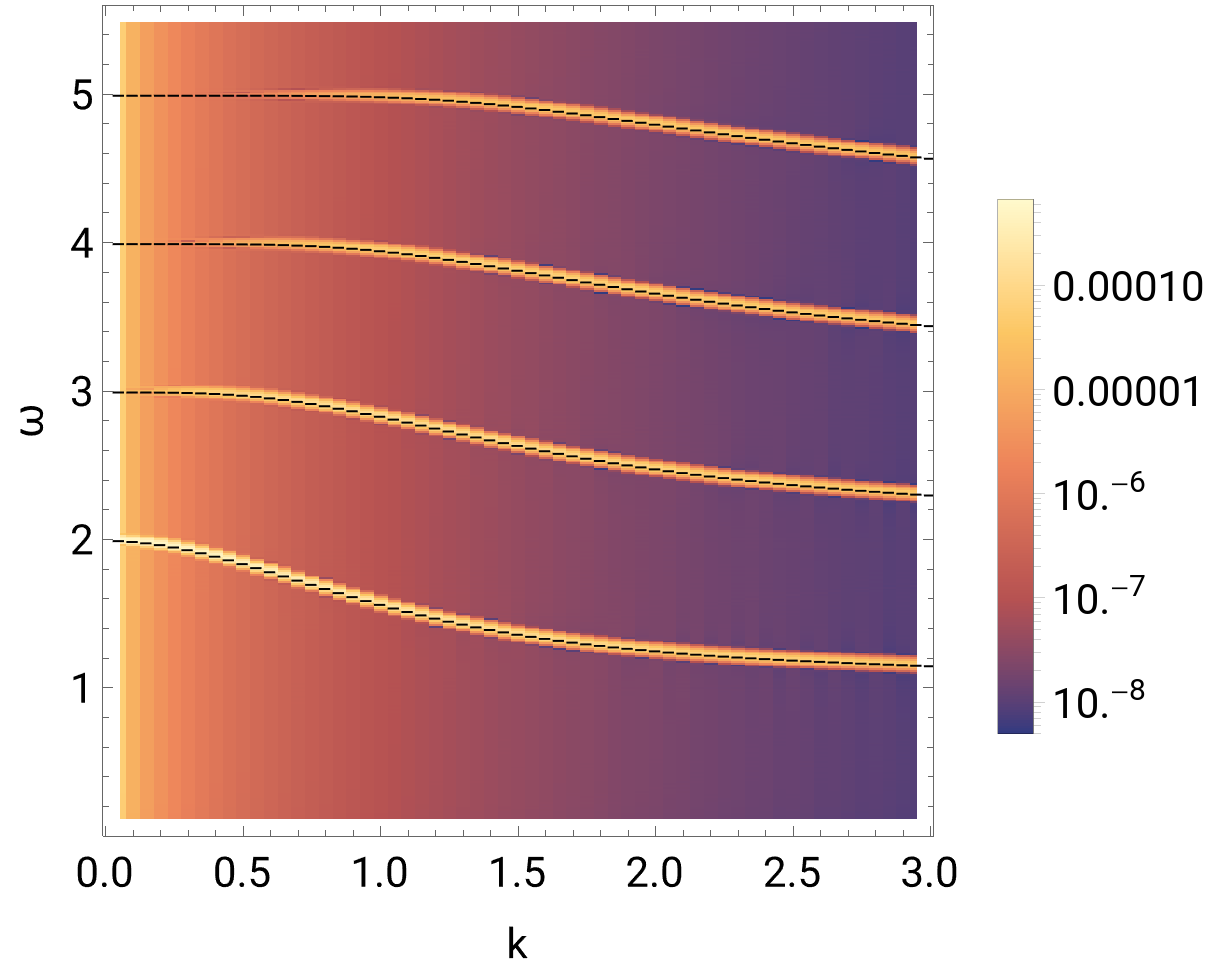}
        \caption{Pure IBW ($k_x \neq 0$).}
        \label{fig:pure_ibw}
    \end{subfigure}
    \caption{Verification of the two Bernstein wave branches. Black dashed lines indicate the roots of the analytical dispersion relations \eqref{neutIBW} and \eqref{pureIBW}, while the color map represents the magnitude of the numerical Fourier amplitude of the electric potential. Simulations were performed on a $512 \times 33 \times 33$ 1D2V grid with domain dimensions $[0, 20 \pi] \times[-6,6] \times [-6,6]$, $\Delta t = 0.02$, $T = 160 \pi$, and $\alpha = 10^{-6}$. To account for the non-periodicity of the time-signal, the data was multiplied by a Gaussian window prior to the Fourier transform, which explains the finite spectral width of the numerical solutions.}
\end{figure}

For $B \neq 0$, the electrostatic dispersion relation derived from the linearized Vlasov equation is given by \cite{brambilla1998, stix}
\begin{equation}
    \label{ibwDispRel}
    1 + \frac{\hat{n}_1}{\hat{\phi}} + \sum_{p=-\infty}^{\infty} \frac{\omega}{\sqrt{2} v_{\text{th}} k_z} Z \left ( \frac{\omega - \Omega p}{\sqrt{2} v_{\text{th}} k_z} \right ) \Gamma_p(\rho^2 k_\perp^2) = 0 
\end{equation}
where
\begin{equation}
    \label{bernsteinGamma}
    \Gamma_p (x) \coloneq e^{-x} I_p(x) .
\end{equation}
Here, $I_p$ is the modified Bessel function of the first kind, $\hat{n}_1$ is the linearized density perturbation relative to the background density $n_0$, $Z$ is the plasma dispersion function \cite{nrl}, $\Omega$ is the ion gyrofrequency, $\rho = m v_{\text{th}}/q B$ is the ion gyroradius, and $k_\perp$ is the wavevector perpendicular to the magnetic field direction. In the limit $k_z \rightarrow 0$, Eq. (\ref{ibwDispRel}) reduces to
\begin{equation}
    \label{ibwDispRelkz0}
    1 + \frac{\hat{n}_1}{\hat{\phi}} - \sum_{p= - \infty}^{\infty} \frac{\omega}{\omega - \Omega p} \Gamma_p(\rho^2 k_\perp^2) = 0 .
\end{equation}
Notably, the solutions of Eq. (\ref{ibwDispRelkz0}) are not subject to Landau damping, thus providing a robust benchmark for code verification. Our field solver distinguishes between two different branches of Bernstein waves based on the orientation of the perturbation relative to the flux-surfaces. For $k_z = 0$, but $k_y \neq 0$, the perturbation varies within the flux-surface, and due to the periodic boundary conditions, $\langle \hat{n}_1 \rangle = \langle \hat{\phi}_1 \rangle = 0$. For small perturbations, the potential follows from the linearized Boltzmann relation (\ref{cfsaPotentialb}),
\begin{equation}
    \hat{\phi} = \frac{T_e}{e} \frac{\hat{n}_1}{n_0} .
\end{equation}
Assuming $T_e = e = n_0 = 1$, Eq. (\ref{ibwDispRel}) gives the "neutralizing" ion Bernstein waves dispersion relation (Fig. \ref{fig:neut_ibw})
\begin{equation}
    \label{neutIBW}
    2 - \sum_{p=-\infty}^{\infty} \frac{\omega}{\omega - \Omega p} \Gamma_p(\rho^2 k_\perp^2) = 0 .
\end{equation}
Conversely, for $k_y = k_z = 0$, but $k_x \neq 0$, the perturbation is strictly orthogonal to the flux surfaces. In this case, the field solver enforces $\hat{n}_1 = \langle \hat{n}_1 \rangle = 0$ to maintain quasi-neutrality, and Eq. (\ref{ibwDispRel}) gives the "pure" ion Bernstein wave dispersion relation (Fig. \ref{fig:pure_ibw}),
\begin{equation}
    \label{pureIBW}
    1 - \sum_{p=-\infty}^{\infty} \frac{\omega}{\omega - \Omega p} \Gamma_p(\rho^2 k_\perp^2) = 0 .
\end{equation}
The numerical results exhibit excellent agreement with the analytical solutions, as illustrated in Figures \ref{fig:neut_ibw} and \ref{fig:pure_ibw}.

\section{Conclusion}
\label{sec:conclusion}

The present BSL6D framework simulates fully kinetic ions with adiabatic electrons, capturing ion-scale phenomena such as Bernstein waves and ion temperature gradient instabilities. However, a comprehensive treatment of edge turbulence requires kinetic electrons, whose direct coupling via Poisson's equation introduces fast waves and severe time-step restrictions. Quasi-neutrality eliminates these modes, but determining the corresponding electric potential is non-trivial in fully kinetic models where the ion polarization is implicitly contained in the distribution function.

In this work, we have presented a novel, implicit method to derive the quasi-neutral electric potential directly from the semi-Lagrangian advection algorithm. We have demonstrated the approach for the massless limit of drift-kinetic electrons under the assumption of a small magnetic shear. In this limit, the flux-surface averaged electron density remains constant, rendering the problem mathematically equivalent to eliminating the Langmuir-like oscillations in the flux-surface averaged ion density.

A central result is the proof that our implicitly determined electric potential preserves the global $O(\Delta t^2)$ convergence of the underlying Strang-splitting scheme. Furthermore, we have carefully considered the impact of interpolation errors and introduced spectral correction mechanisms to robustly maintain the quasi-neutrality constraint. Without these corrections, unphysical density drifts would occur in the presence of large background density variations typical of the plasma edge region, ultimately leading to numerical instabilities.

We supplemented our theoretical derivations with rigorous numerical verifications. Most notably, we have demonstrated the second-order error convergence of the electric potential and successfully verified two branches of Bernstein waves. Extending this implicit quasi-neutral field solver to accommodate massive, drift-kinetic electrons is the primary objective of future work, where the current electrostatic framework will serve as a thoroughly verified baseline for advanced iterative solvers. This will ultimately enable self-consistent simulations of drift-kinetic electrons in tokamak edge plasmas, where the interplay of kinetic ion and electron dynamics governs critical confinement transitions such as the L-H transition.

\section{Acknowledgments}
    This work has been carried out partly within the framework of the EUROfusion
    Consortium, funded by the European Union via the Euratom Research and Training
    Programme (Grant Agreement No 101052200 – EUROfusion). Support has also been
    received by the EUROfusion High Performance Computer (Marconi-Fusion). Views and
    opinions expressed are however those of the author(s) only and do not
    necessarily reflect those of the European Union or the European Commission.
    Neither the European Union nor the European Commission can be held responsible
    for them.  Numerical simulations were performed at the MARCONI-Fusion
    supercomputer at CINECA, Italy, and at the HPC system at the Max Planck
    Computing and Data Facility (MPCDF), Germany.

\bibliographystyle{unsrt}
\bibliography{references}

\end{document}